\documentclass[reprint,twocolumn,amsmath,amssymb,aps,prl]{revtex4-1}
\usepackage{graphicx,graphics,color,epsfig}
\usepackage{bm}
\usepackage{amsmath}
\usepackage{mathrsfs}
\usepackage{amssymb}
\usepackage{subfigure}
\usepackage{amsfonts}
\usepackage{graphicx}
\usepackage{multirow,booktabs}
\usepackage{epstopdf}
\usepackage{diagbox}
\usepackage[colorlinks,
anchorcolor=blue,  
linkcolor=blue,       
citecolor=blue]{hyperref}  
\usepackage{natbib}

\newcommand{\be}{\begin{eqnarray}}
\newcommand{\ee}{\end{eqnarray}}

\newcommand{\ba}{\begin{align}}
\newcommand{\ea}{\end{align}}

\def\sgn{\mathop{\rm sgn}}

\providecommand{\U}[1]{\protect\rule{.1in}{.1in}}
\begin{document}
\title{Impurity-driven transitions in frustrated quantum Ising ring}
\author{Han-Chuan Kou}
\affiliation{College of Physics, Sichuan University, 610064, Chengdu, People’s Republic of China\\
and Key Laboratory of High Energy Density Physics and Technology of Ministry of Education, Sichuan University, 610064,
Chengdu, People’s Republic of China}
\author{Zhen-Yu Zheng}
\affiliation{College of Physics, Sichuan University, 610064, Chengdu, People’s Republic of China\\
and Key Laboratory of High Energy Density Physics and Technology of Ministry of Education, Sichuan University, 610064,
Chengdu, People’s Republic of China}
\author{Peng Li}
\email{lipeng@scu.edu.cn}
\affiliation{College of Physics, Sichuan University, 610064, Chengdu, People’s Republic of China\\
and Key Laboratory of High Energy Density Physics and Technology of Ministry of Education, Sichuan University, 610064,
Chengdu, People’s Republic of China}

\date{\today}

\begin{abstract}
  We study the quantum phase transitions driven by a point impurity in a chain seamed with ring frustration. Rich phases and quantum phase transitions are uncovered and characterized by both bulk and impurity correlation functions. Nonlocality of the correlation functions are emphasized in manifesting the novel features in the system. We demonstrate that the long-range correlation function can be factorized into local and nonlocal factors in the thermodynamic limit. The gapless topological extended-kink (TEK) phase is disclosed to exhibit long-range correlation but without long-range order, because its ground state is nondegenerate and thus immune to spontaneous symmetry breaking. This conclusion is also true in the classical impurity limit, which is significantly different from that for the open boundary chain without ring frustration. However, spontaneous symmetry breaking does occur in the gapped kink zero mode (KZM) phase and leads to the antiferromagnetic zero mode (AFZM), in which antiferromagnetic order develops in the bulk while entangled states persists locally around the impurity. And as a new feature of quantum phase transition induced by impurity, the transition from the TEK phase to the KZM-AFZM phase is reflected by a steplike nonlocal factor of the correlation function.
\end{abstract}

\maketitle


\section{Introduction}
In the field of quantum phase transition \cite{sachdev_2011}, impurity can play important roles, for instance, it can induce remarkable bulk effects in critical or quasicritical systems \cite{Jacobsen_2008} and pave the way to design quantum devices \cite{PRA_Lorenzo_2013, PRA_Lorenzo_2015}. On the other hand, geometrical spin frustration in low dimensions can induce strong quantum fluctuations that lead to interesting phenomena \cite{Diep_2005}. Recently, the effect of ring frustration becomes attraction because it can provide robust exotic low-energy states that maintain quantum coherence \cite{SovPJ_Bariev_1979, PRL_Jullien_1986, PRB_Jullien_1987, AdvancesP_Igloi_1993, PRE_Vicari_2015, JSM_Li_2016, PRE_Li_2018, PRE_Li_2019, JPC_Franchini_2019}. The systems with periodic or antiperiodic boundary conditions (PBC or APBC) are of great theoretical interests because of the fascinating phase transitions and critical phenomena in them \cite{Francesco_2012, Milsted_PRB_2017, Zou_PRL_2018, PRL_Zou_2020, PRB_Zou_2020}. Attention also arises due to impressive progresses in designing and fabricating quantum devices with ring structure for achieving purposed applications \cite{Nature_Labuhn_2016, PRL_Jin_2019}.

To explore the joint effect of the impurity and ring frustration is an intriguing topic. In systems with ring frustration, bond impurity (or bond defect) has been introduced and studied \cite{PRE_Vicari_2015, PRB_Li_2019, arxiv_Franchini_2020}. In this work, we investigate a point impurity in the quantum Ising chain seamed with ring frustration. The merit of point impurity is that it is easier to control \cite{PRB_Plastina_2016, PRB_Plastina_2017}. The intriguing interplay between the ring frustration and the point impurity leads to a rich ground-state phase diagram. Quantum phase transition induced by the point impurity is characterized by the scaling behavior of bulk and impurity correlation functions and correlation lengths.

The contents are organized as follows: In Sec. II, we go into some details of the rigorous solution of the model, since the quantum Ising chain with PBC is quite different from the one with open boundary condition (OBC) and delicate mapping between the spins and fermions must be looked after.
In Sec. III, we construct the ground-state phase diagram basing on the rigorous solution and the complementary perturbative theory. The latter can give us a simplified picture of the low-energy states.
In Sec. IV, we characterize in detail the new features of the phases and transitions by appropriate correlation functions in the context of nonlocality.
In Sec. V, we give a brief summary.

\section{The model with ring frustration and impurity}

\subsection{The model}

The simplest model containing both ring frustration and impurity reads,
\begin{align}\label{app-2-1-1}
	H=J\sum_{j=1}^{N}\sigma_{j}^{x}\sigma_{j+1}^{x}-h \sum_{j=1}^{N-1}\sigma_{j}^{z}-\mu h \sigma_{N}^{z},
\end{align}
where $\sigma_{j}^{a} (a=x,z)$ are Pauli matrices, PBC is adopted, i.e. $\sigma_{N+j}^{a}=\sigma_{j}^{a}$. The geometrical ring frustration in the first term is guaranteed by the odd total number of lattice sites, $N\in \mathrm{odd}$, and the antiferromagnetic coupling, $J>0$. We shall set the reference energy scale, $J=1$, henceforth. The transverse fields are tunable so as to realize a \emph{heavier} point impurity at site $N$ for $\mu>1$ and a \emph{lighter} one for $0<\mu<1$.

By the Jordan-Wigner transformation,
\be
  &f_{j}^{\dag}=\frac{1}{2}(\sigma_{j}^{x}+i \sigma_{j}^{y})\prod_{l=1}^{j-1}(-\sigma_{l}^{z}), (1\leq j\leq N), \label{JWT}
\ee
one can find that the exact solution of $H_{\mathrm{P}}$ comes from the two free-fermion Hamiltonians,
\begin{align}
  H^{\mathrm{R/NS}}=&\sum_{j=1}^{N}(f_{j}^{\dagger}f_{j+1}+f_{j+1}f_{j} + H.c.) \nonumber\\
    &-h \sum_{j=1}^{N-1}(2f_{j}^{\dagger}f_{j}-1)-\mu h (2f_{N}^{\dagger}f_{N}-1),
\end{align}
where the superscript $\mathrm{R/NS}$ means the ``Ramond" sector or PBC, $f_{N+1}=f_1$, and the ``Neveu-Schwarz" sector or APBC, $f_{N+1}=-f_{1}$, respectively. The solution of the aimed Hamiltonian $H$ is obtained by the projection,
\begin{align}
  H = P_{z}^{-} H^{\mathrm{R}} + P_{z}^{+} H^{\mathrm{NS}}, \label{H_P}
\end{align}
where the projectors
\be
  P_{z}^{\pm}&=&\frac{1}{2}(1\pm\mathscr{P}_{z})
\ee
are defined based on the parity operator
\be
  \mathscr{P}_{z} &=& \prod_{j=1}^{N} (-\sigma_{j}^{z}).
\ee
The parity operator commutes with the Hamiltonian $H$, which facilitates us to solve the system by tedious but clear projections.

It is noteworthy that the redundant degrees of freedom in $H^{\mathrm{R}}$ and $H^{\mathrm{NS}}$ compose another quantum Ising ring with APBC, $\tilde{H} = P_{z}^{+} H^{\mathrm{R}} + P_{z}^{-} H^{\mathrm{NS}}$. So that the four relevant Hamiltonians are linked together by a complete \emph{quaternary Jordan-Wigner mapping} \cite{PRB_Li_2019}.

\begin{figure}[t]
  \begin{center}
  \includegraphics[width=3.5in,angle=0]{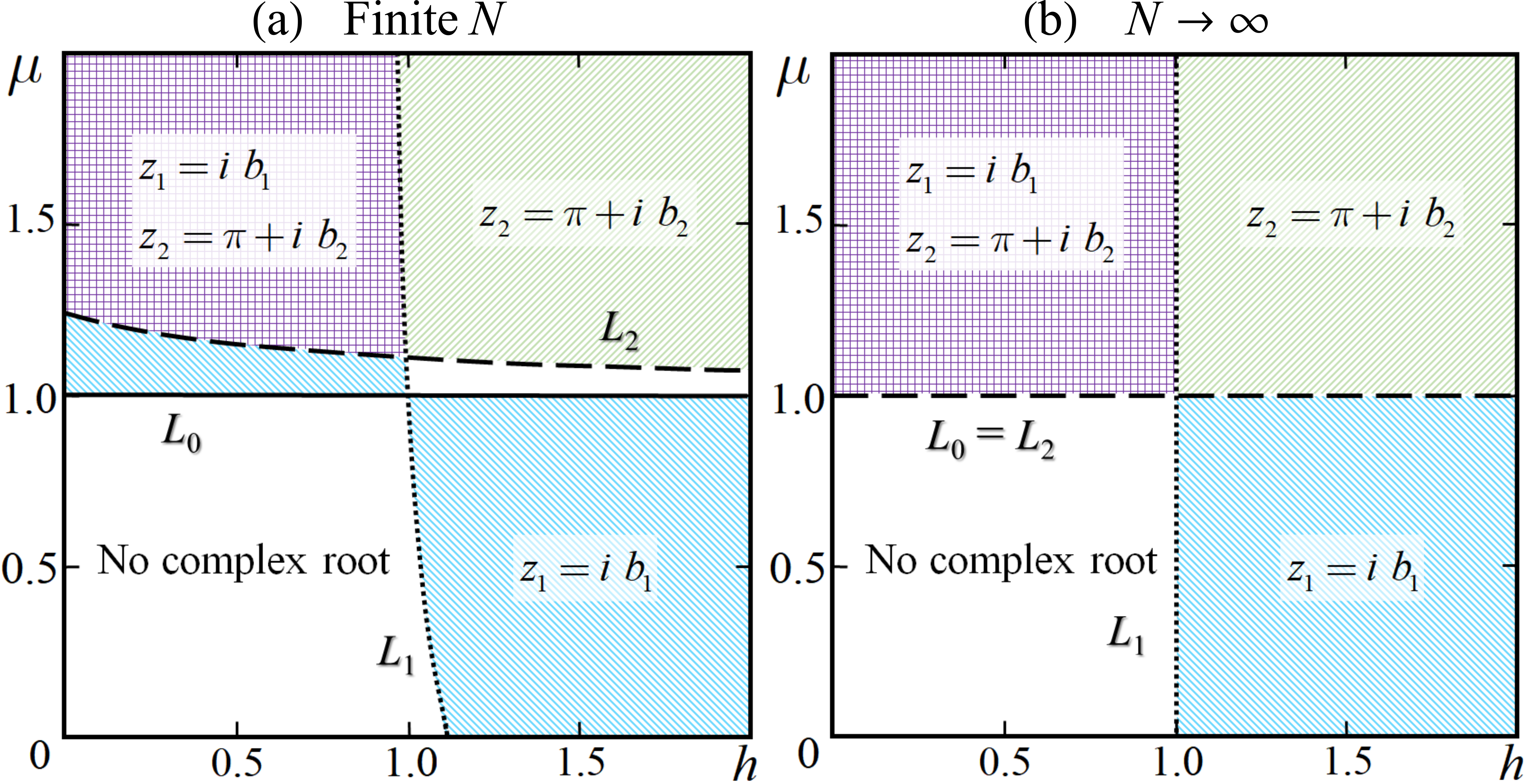}
  \end{center}
  \caption{Distribution of the complex roots for $H_{\mathrm{R}}$ in the parameter plane $(h,\mu)$ for: (a) finite $N$ (Here, we take $N=9$ for demonstration); (b) the thermodynamic limit, $N\rightarrow\infty$. Please see more details in the text.}%
  \label{rootz1z2}%
\end{figure}

\subsection{Solution}

Because this tedious mapping is different from that in the case of open boundary condition (OBC) \cite{PRB_Plastina_2017}, we give some details about the solution of the system. The Hamiltonians, $H^{\mathrm{R/NS}}$, are solved according to the procedure originally stated by Lieb \emph{et al} \cite{Annals_Lieb_1961, EPJB_Igloi_1988}. It resorts to finding out the fermionic quasiparticle operators,
\be
  \eta_{q}=\sum_{j=1}^{N}(g_{q,j}f_{j}+h_{q,j}f_{j}^{\dagger}), \\
  \eta_{q}^{\dagger}=\sum_{j=1}^{N}(g_{q,j}f_{j}^{\dagger}+h_{q,j}f_{j})
\ee
The solution of coefficients are exposed in Appendix \ref{Appendix_A}. Now we explain some details in the odd channel $P_{z}^{-} H_{\mathrm{R}}$ first, since it provides the ground state for arbitrary $N$. After diagonalization, we arrive at
\be
  H^{\mathrm{R}}=\sum_{z}\Omega_{z} \eta_{z}^{\dagger} \eta_{z} +\sum_{q}\mathrm{'}\omega(q)\eta_{q}^{\dagger}\eta_{q},
\ee
where
\begin{equation}
  \omega(q) = 2 \sqrt{1+h^{2} - 2 h \cos q}.
\end{equation}
and the possible discrete mode(s), $\Omega_{z}$, are singled out and excluded in the sum $\sum_{q}\mathrm{'}$ if they appear.

\begin{figure}[t]
  \begin{center}
  \includegraphics[width=3.5in,angle=0]{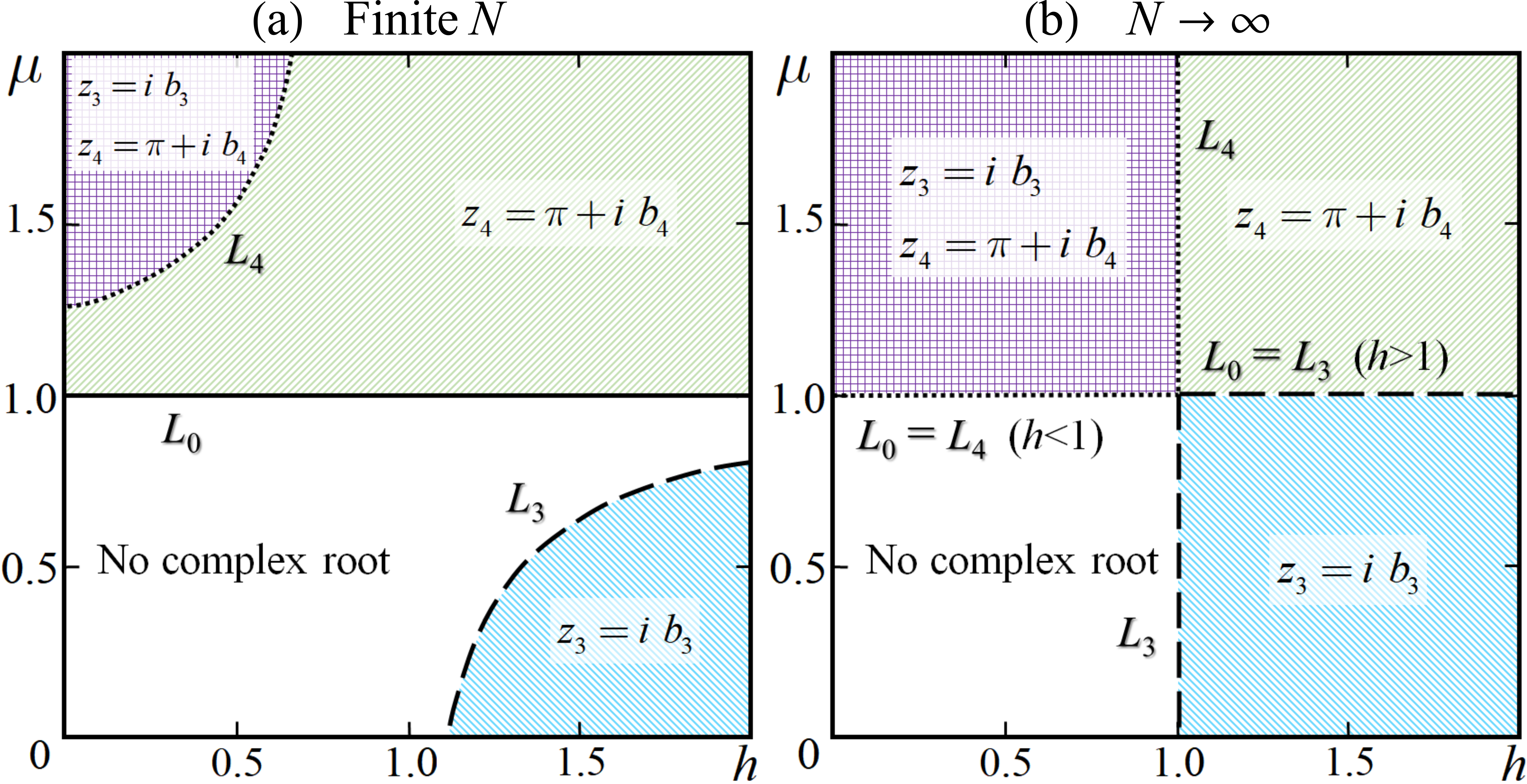}
  \end{center}
  \caption{Distribution of the complex roots for $H_{\mathrm{NS}}$ in the parameter plane $(h,\mu)$ for: (a) finite $N$ (Here, we take $N=9$ for demonstration); (b) the thermodynamic limit, $N\rightarrow\infty$. Please see more details in the text.}%
  \label{rootz3z4}%
\end{figure}

The values of $N$ independent $q$ (including possible complex $z$) are roots of the equation,
\begin{align}
  2&\mathscr{P}_{z}\mu \sin q + h (\mu^{2}-1) \sin Nq \nonumber\\
  &+ \sin(N+1)q-\mu^{2} \sin(N-1)q = 0, \label{Lieb_eqn}
\end{align}
where $\mathscr{P}_{z}=-1$ and $+1$ for $H^{\mathrm{R}}$ and $H^{\mathrm{NS}}$ respectively. Now we explain the case for $H^{\mathrm{R}}$ first. We start from the translationally symmetric case, $\mu=1$, where the roots of $q$ are $N$ real commensurate values \cite{JSM_Li_2016},
\begin{equation}
  \{-\frac{N-1}{N}\pi,\cdots,-\frac{2}{N}\pi,0,\frac{2}{N}\pi,\cdots,\frac{N-1}{N}\pi\}.
  \label{q_values}
\end{equation}
When deviating from the symmetric case, the values of $q$ depart from the commensurate ones. Furthermore we may get complex root $z = a + i b$ reflecting the discrete modes. There are four situations for the appearance of the complex roots:

(i) Two complex roots and $(N-2)$ real roots. We label the complex roots as
\begin{align}
  z_{1} &= i b_{1}, \label{z1}\\
  z_{2} &= \pi + i b_{2}, \label{z2}
\end{align}
where $b_{1}$ and $b_{2}$ are real numbers.

(ii) One complex root, $z_{1}= i b_{1}$, and $(N-1)$ real roots.

(iii) One complex root, $z_{2}= \pi + i b_{2}$, and $(N-1)$ real roots.

(iv) $N$ real roots.

\noindent In all situations, the possible complex roots change to incommensurate real values continuously,
\be
  z_{1} \rightarrow 0 \rightarrow a_{1}, \\
  z_{2} \rightarrow \pi \rightarrow a_{2},
\ee
when we go across the dividing lines in the parameter plane $(h,\mu)$ as illustrated in Fig. \ref{rootz1z2},
\begin{align}
  &L_{0}:~~\mu=1,\\
  &L_{1}:~~\mu=\frac{1+N-N h}{1-N+N h},\\
  &L_{2}:~~\mu=\frac{1+N+N h}{-1+N+N h}.
\end{align}
In the thermodynamic limit $N\rightarrow\infty$, $L_{1}$ becomes a vertical line $h=1$ and $L_{2}$ approaches $L_{0}$ asymptotically with a difference of the order $O(1/N)$ (Fig. \ref{rootz1z2}(b)).

The solution for the even channel $P_{z}^{+} H^{\mathrm{NS}}$ is similar. However, the commensurate values for the roots of $q$ are,
\begin{equation}
  \{-\frac{N-2}{N}\pi,\cdots,-\frac{1}{N}\pi,\frac{1}{N}\pi,\cdots,\frac{N-2}{N}\pi, \pi\}.
\end{equation}
The possible complex roots are labelled by
\be
  z_{3} &=& i b_{3}, \label{z3}\\
  z_{4} &=& \pi + i b_{4}.
\ee
They also change to incommensurate real values continuously,
\be
  z_{3} \rightarrow 0 \rightarrow a_{3}, \\
  z_{4} \rightarrow \pi \rightarrow a_{4},
\ee
when we go across the dividing lines as shown in Fig. \ref{rootz3z4},
\begin{align}
  &L_{0}:~~\mu=1,\\
  &L_{3}:~~\mu=\frac{-1-N+N h}{1-N+N h},~~~(h>1),\\
  &L_{4}:~~\mu=\frac{-1-N+N h}{1-N+N h},~~~(h<1).
\end{align}
In the thermodynamic limit, $L_{3}$ and $L_{4}$ asymptotically approaches the vertical line $h=1$ and horizontal line $\mu=1$ in the order $O(1/N)$ with $N\rightarrow\infty$ as shown in Fig. \ref{rootz1z2}(b).

Let the vacua devoid of quasiparticles for $H^{\mathrm{R}}$ and $H^{\mathrm{NS}}$ be denoted by $|0^{\mathrm{R}}\rangle$ and $|0^{\mathrm{NS}}\rangle$ respectively. Then we can recover the valid states of the aimed Hamiltonian $H$ by picking out the valid states in the odd channel, $P_{z}^{-} H^{\mathrm{R}}$, and the even channel, $P_{z}^{+} H^{\mathrm{NS}}$.

\section{Ground-state phase diagram}

\begin{figure}[t]
  \includegraphics[width=3.0in,angle=0]{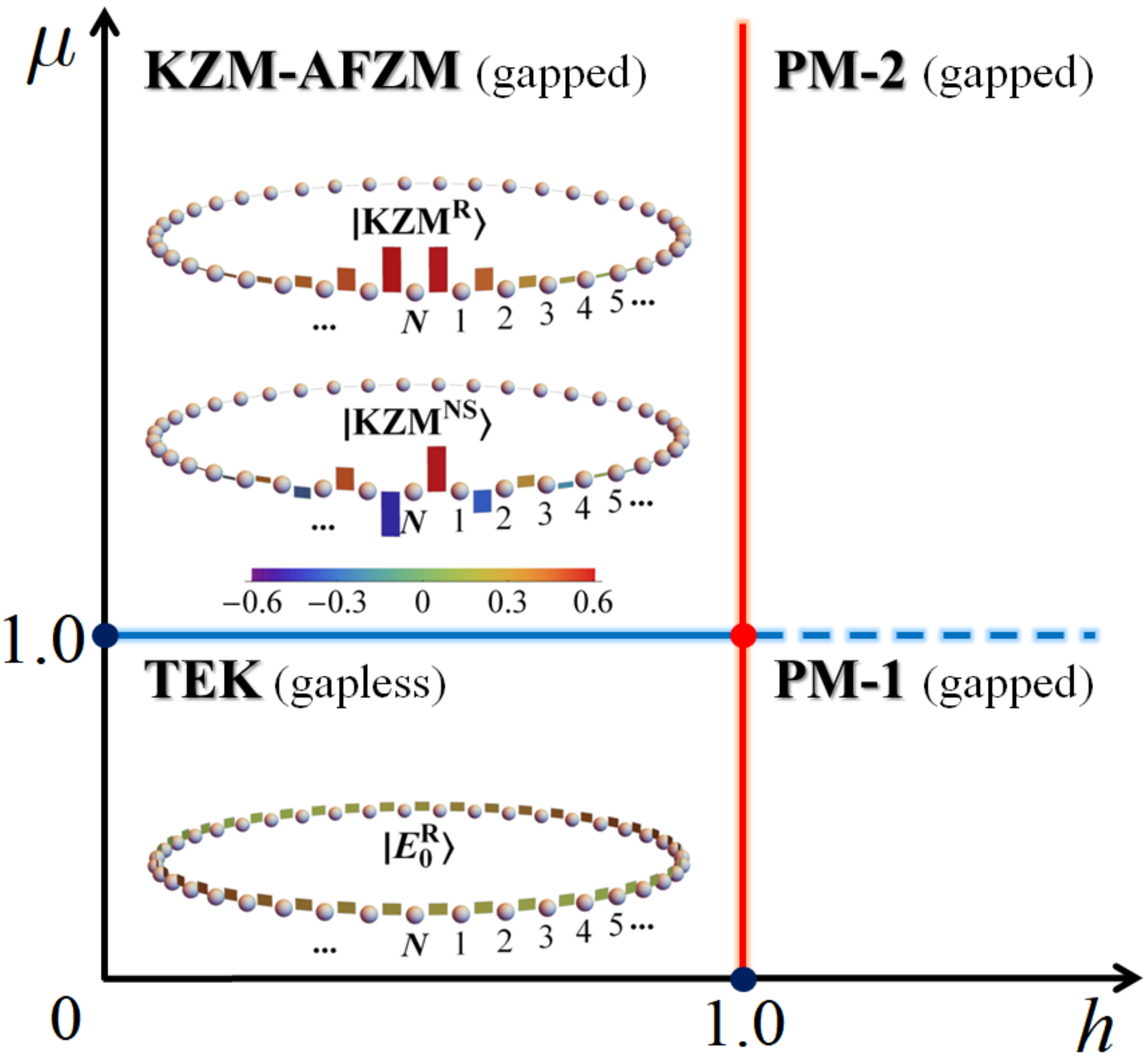}
  \caption{Ground-state phase diagram. The ground state is doubly degenerate in the KZM-AFZM phase, nondegenerate in the gapless TEK and gapped PM phases (with two subphases, PM-1 and PM-2). The depicted ground states, $|E_{0}^{\mathrm{R}}\rangle$ with parameters $(h,\mu)=(0.1,1)$ in the TEK phase and $|\mathrm{KZM}^{\mathrm{R/NS}}\rangle$ with parameters $(h,\mu)=(0.1,2)$ in the KZM-AFZM phase, are obtained by perturbative treatment on a system with $N=41$ for demonstration.}%
  \label{lowlevels}%
\end{figure}

In the thermodynamic limit, $N\rightarrow\infty$, we obtain the ground-state phase diagram, Fig. \ref{lowlevels}, containing three phases. They are: (i) gapless topological extended-kink (TEK) phase ($\mu<1$ and $h<1$); (ii) gapped kink zero mode (KZM) phase ($\mu>1$ and $h<1$); (iii) gapped paramagnetic (PM) phase ($h>1$). The translationally symmetric line, $\mu=1$ and $h<1$, belongs to the TEK which has been disclosed previously \cite{JSM_Li_2016,PRB_Li_2019}. It is topological in the sense that we can work out its nontrivial winding number, $w=1$ \cite{PRB_Li_2019}. In the gapless TEK, $2N$ quantum energy states compose the lowest band of width $4h$. The band is quasicontinuous since the differences of the energy levels are in the order $O(1/N)$. The $\mathscr{P}_{z}$ symmetry of KZM can be broken, which leads to the antiferromagnetic zero mode (AFZM) that will be disclosed later. So we label it as the KZM-AFZM phase. Divided by the line $\mu = 1$, the PM-1 and PM-2 subphases are distinct according to the first excited state.

\begin{figure}[t]
  \includegraphics[width=3.5in,angle=0]{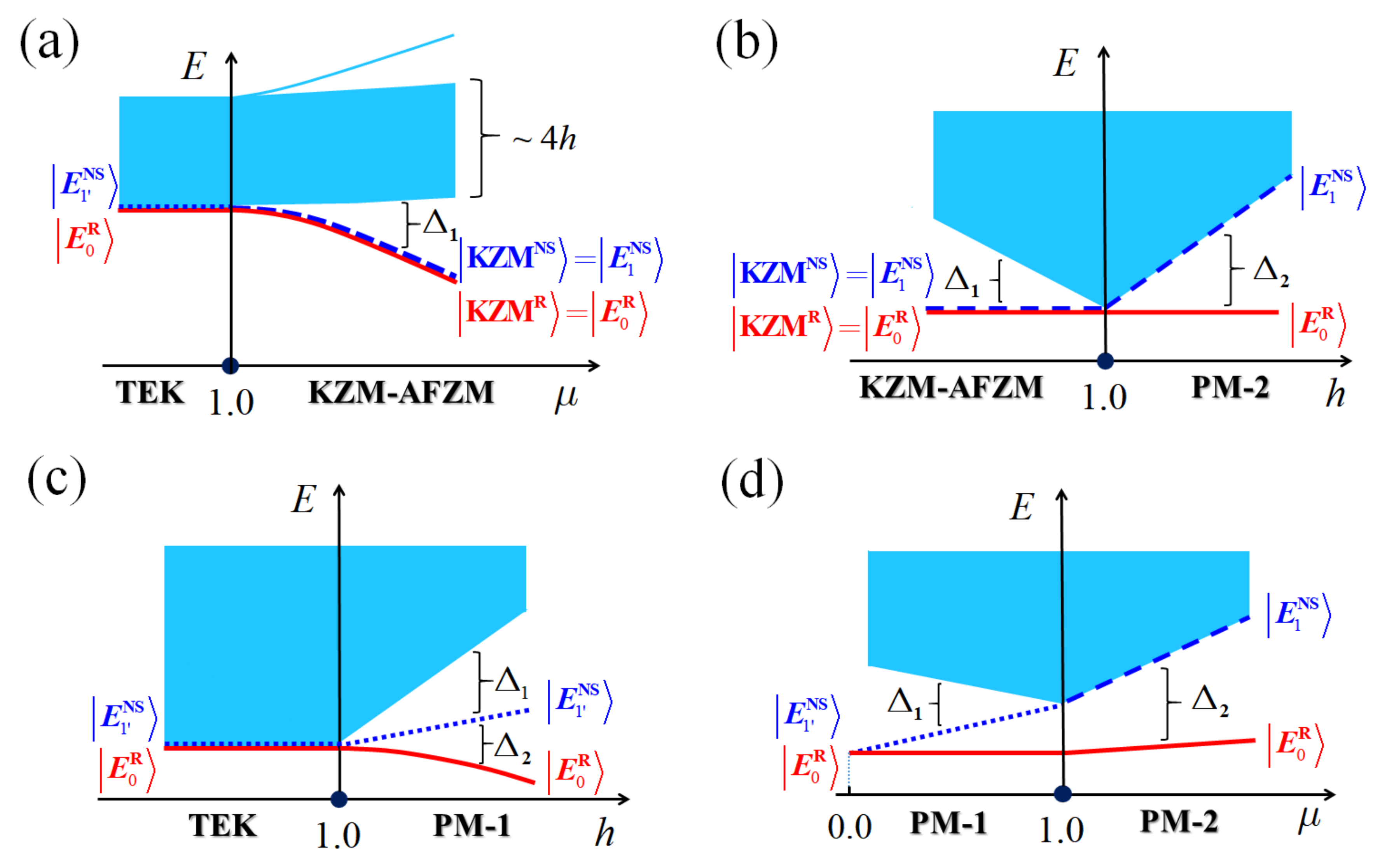}
  \caption{Schematics of the low-energy band and levels showing the transitions: (a) from TEK to KZM-AFZM; (b) from KZM-AFZM to PM-2; (c) from TEK to PM-1; (d) from PM-1 to PM-2.}%
  \label{schematics}%
\end{figure}

We can observe the phases by drawing the lowest energy band and levels as sketched in Fig. \ref{schematics}.

For both finite and infinite $N$, the ground state
\be
  |E_{0}^{\mathrm{R}}\rangle = \eta_{z_{1}}^\dagger|0^{\mathrm{R}}\rangle, \label{E0}
\ee
evolves adiabatically in all phases. The superscript R means that the state comes from the odd channel $P_{z}^{-} H^{\mathrm{R}}$.

The first excited state comes from the even channel $P_{z}^{+} H^{\mathrm{NS}}$ and can be expressed as,
\be
  |E_{1}^{\mathrm{NS}}\rangle = \eta_{z_{3}}^\dagger\eta_{z_{4}}^\dagger|0^{\mathrm{NS}}\rangle. \label{E_1}
\ee
However, it only evolves adiabatically from TEK to PM-1 and from KZM-AFZM to PM-E, so we use Eq. (\ref{E_1}) in KZM-AFZM and PM-2 and use a distinct  notation $|E_{1'}^{\mathrm{NS}}\rangle$ in TEK and PM-1 (Fig. \ref{schematics}). It evolves non-adiabatically from TEK to KZM-AFZM and from PM-1 to PM-2 due to energy level crossing at the boarder. We define $\Delta_1$ as the gap between $|E_{1}^{\mathrm{NS}}\rangle/|E_{1'}^{\mathrm{NS}}\rangle$ and above continuous band, and $\Delta_2$ as the gap between $|E_{1}^{\mathrm{NS}}\rangle/|E_{1'}^{\mathrm{NS}}\rangle$ and $|E_{0}^{\mathrm{R}}\rangle$.

In KZM-AFZM, the gap $\Delta_2$ becomes zero (in the order $O(e^{-N})$), so we get two degenerate KZM states,
\be
  &&|\mathrm{KZM}^{\mathrm{R}}\rangle = |E_{0}^{\mathrm{R}}\rangle, \label{KZM-R}\\
  &&|\mathrm{KZM}^{\mathrm{NS}}\rangle = |E_{1}^{\mathrm{NS}}\rangle. \label{KZM-NS}
\ee
Above them, the gap $\Delta_{1}$ develops,
\be
  \Delta_{1} = 2(1-h)-2\sqrt{1+h^{2}-2h\cos z_{1}},
\ee
where
\be
  z_{1} = i \ln\frac{h(1-\mu^{2})+\sqrt{4\mu^{2}+h^{2}(1-\mu^{2})^{2}}}{2}  \label{z1_exa}
\ee
is an exact solution of  Eq. (\ref{Lieb_eqn}) in the limit $N\rightarrow\infty$ as described in Eq. (\ref{z1}). The two KZM states can be best visualized by a perturbative treatment that is a beneficial supplement to the rigorous solution for TEK and KZM-AFZM.
The key point is to utilize the lowest $2N$ classical Ising kink states $(j=1,2,\cdots, N)$,
\begin{align}
  |j,\rightarrow\rangle = |\cdots,\leftarrow_{j-1},\boxed{\rightarrow_{j},\rightarrow_{j+1}},\leftarrow_{j+2},\cdots\rangle,\label{jright}\\
  |j,\leftarrow\rangle = |\cdots,\rightarrow_{j-1},\boxed{\leftarrow_{j},\leftarrow_{j+1}},\rightarrow_{j+2},\cdots\rangle, \label{jleft}
\end{align}
of the frustrated Ising Hamiltonian,
\be
  H_0=\sum_{j=1}^{N}\sigma_{j}^{x}\sigma_{j+1}^{x},
\ee
and take the rest part of the Hamiltonian,
\be
  V = H - H_0,
\ee
as a perturbation ($h\ll 1$), which leads to entangled kink states as the eigenstates. For example, the ground state in the symmetric TEK ($\mu=1$) approximately reads,
\be
  |E_{0}^{\mathrm{R}}\rangle \approx \frac{1}{\sqrt{2N}}\sum_{j=1}^{N}(|j,\rightarrow\rangle+|j,\leftarrow\rangle).
\ee
While the two KZM states can be expressed as
\be
  |\mathrm{KZM}^{\mathrm{R}}\rangle \approx \sum_{j=1}^{N}\psi_{j}(|j,\rightarrow\rangle + |j,\leftarrow\rangle),\label{approxKZM_R}\\
  |\mathrm{KZM}^{\mathrm{NS}}\rangle \approx \sum_{j=1}^{N}\chi_{j}(|j,\rightarrow\rangle - |j,\leftarrow\rangle).\label{approxKZM_NS}
\ee
The coefficients are depicted in the inset of Fig. \ref{lowlevels}, which shows a localization behavior of the KZM states. Detailed expressions can be found in Appendix \ref{Appendix_B}. The $\mathscr{P}_{z}$ symmetry breaking of KZM and the formation of AFZM will be elaborated later.

\section{Long-range and short-range correlations in the phases and transitions}

Now we resort to the appropriate correlation functions in disclosing their intriguing properties and transitions between each other and among others, because the length scales related to the bulk and impurity are embedded in them. The TEK and KZM-AFZM phases are the main focus. Two complementary methods, the perturbative theory and rigorous FSS analysis, are employed.

\subsection{Definitions of correlation functions}

\begin{figure}[t]
	\begin{center}
		\includegraphics[width=1.5in,angle=0]{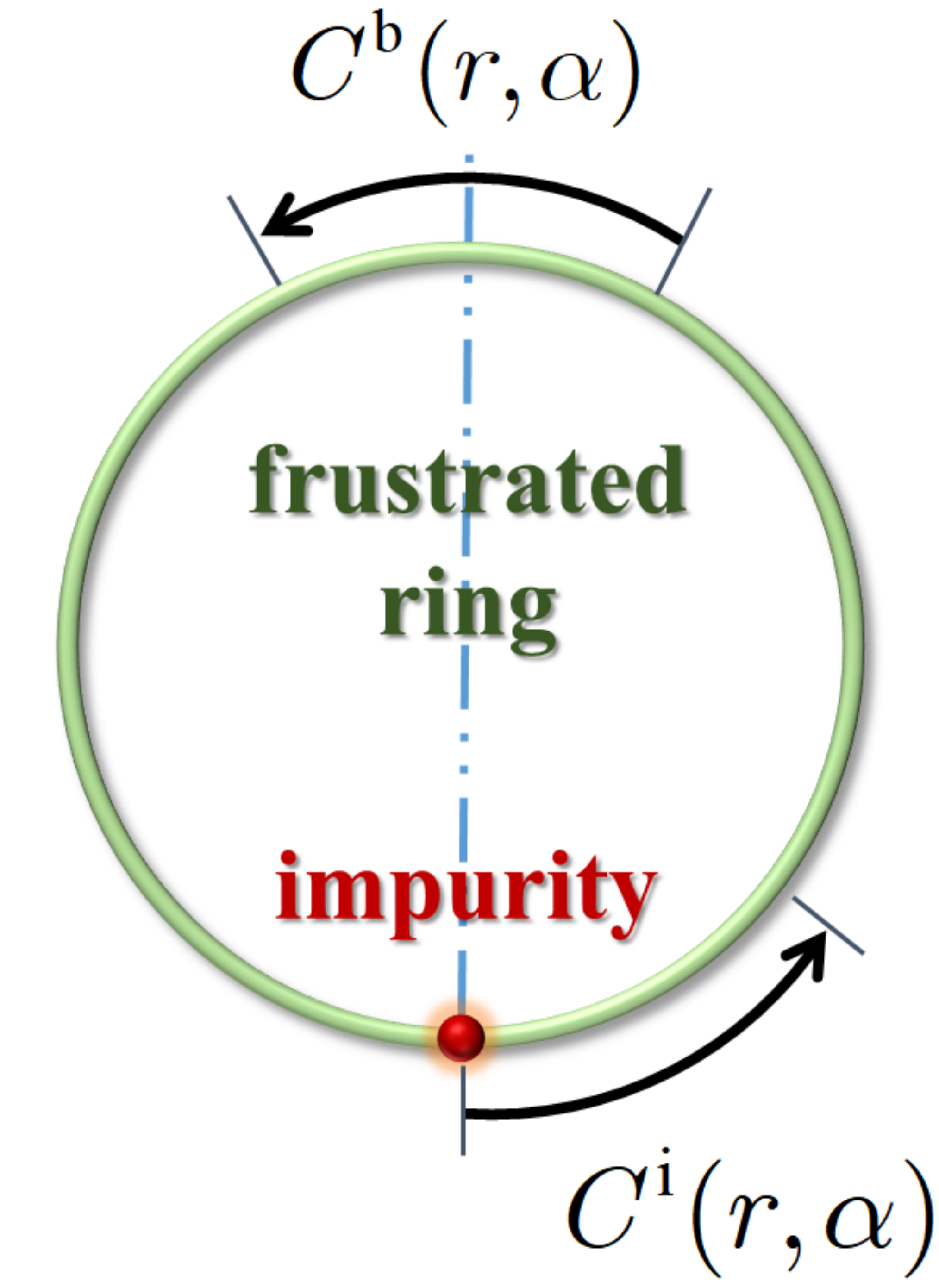}
	\end{center}
	\caption{Two types of two-site longitudinal correlation functions, $C^{\mathrm{b}}(r,\alpha)$ and $C^{\mathrm{i}}(r,\alpha)$, for bulk and impurity. Notice that the two sites in $C^{\mathrm{b}}(r,\alpha)$ are symmetric about the impurity.}%
	\label{ring}%
\end{figure}

\subsubsection{Two-site correlation}

The two-site longitudinal correlation function between site $j$ and $j+r$ for the ground state is defined as
\begin{align}
  C_{j,j+r}^{xx}=\langle \sigma_{j}^{x}\sigma_{j+r}^{x} \rangle, \label{Cxx}
\end{align}
where $\langle\cdots\rangle$ means $\langle E_{0}^{\mathrm{R}}|\cdots| E_{0}^{\mathrm{R}} \rangle$. Throughout the whole paper, we only consider the longitudinal correlation function in $x$ direction. To ease the notation, we will drop the superscript and simply denote it as $C_{j,j+r}$.

By Wick's theorem, the correlation functions can be expressed in determinants as shown in Appendix \ref{Appendix_B}. Although translational symmetry is now broken due to the impurity at site $N$, we have a reflection symmetry instead,
\be
  C_{N-j-r,N-j}=C_{j,j+r}.
\ee

We investigate two types of correlation functions. The first one is the \emph{bulk correlation},
\begin{align}
  C^{\mathrm{b}}_{r,N} \equiv C_{j_{0},j_{0}+r},
\end{align}
where
\be
  j_{0}=\frac{N-1}{2}-\left[\frac{r}{2}\right], (j_{0}\neq N, j_{0}+r\neq N), \label{j0}
\ee
and $\left[\frac{r}{2}\right]$ means taking the integer part of $\frac{r}{2}$. Please notice that the two sites in $C^{\mathrm{b}}_{r,N}$ are symmetric about the impurity (Fig. \ref{ring}). Likewise, the second one is the \emph{impurity correlation},
\begin{align}
  C^{\mathrm{i}}_{r,N} \equiv C_{N,r},
\end{align}
which measures the spin fluctuations between the impurity and another site $r$ in the bulk.

In a system with $N\rightarrow\infty$, we suppose that the measurement of the two-site correlation function can be carried out in a nonlocal distance, $r\rightarrow\infty$, so that a context of nonlocality can be established by defining a scale variable \cite{PRE_Li_2019},
\be
  \alpha =\lim_{N\rightarrow\infty}\frac{r}{N}. \label{alpha}
\ee
Then if $\alpha\neq 0$, we are dealing with a \emph{nonlocal distance} $r$, while if $\alpha=0$ although $r\gg 1$, we get a \emph{local distance}  $r$. Here, we only concern $N\in \mathrm{odd}$. Under this context, the two types of correlations in the thermodynamic limit are denoted by
\begin{align}
  C^{\mathrm{b}}(r,\alpha) \equiv \lim_{N\rightarrow\infty}C^{\mathrm{b}}_{r,N},\\
  C^{\mathrm{i}}(r,\alpha) \equiv \lim_{N\rightarrow\infty}C^{\mathrm{i}}_{r,N}.
\end{align}
As demonstrated by several exactly solvable models, this protocol facilitates us to cope with factorizable correlation in a critical system or a system with long-range correlation (LRC) and is coincident with the numerical finite-size scaling (FSS) analysis \cite{PRE_Li_2019, TANG_PLA_2020}. In this work, we demonstrate the factorizable bulk and impurity correlations and utilize them to characterize the phases and transitions induced by the impurity.

\subsubsection{Three-site correlation}

We define a special three-site correlation function among sites $j$, $j+r$, and $N$ for the ground state,
\begin{align}
  T_{j,j+r,N}=\langle \sigma_{j}^{x}\sigma_{j+r}^{x} \sigma_{N}^{z} \rangle. \label{Tzxx}
\end{align}
$T_{j,j+r,N}$ is ready to be expressed in determinants (Appendix \ref{Appendix_C}) that can be evaluated efficiently for quite large systems. It is an adequate quantity for observing the symmetry breaking of KZM and the formation of AFZM.

\subsection{TEK phase}
\begin{figure}[t]
  \includegraphics[width=3.4in,angle=0]{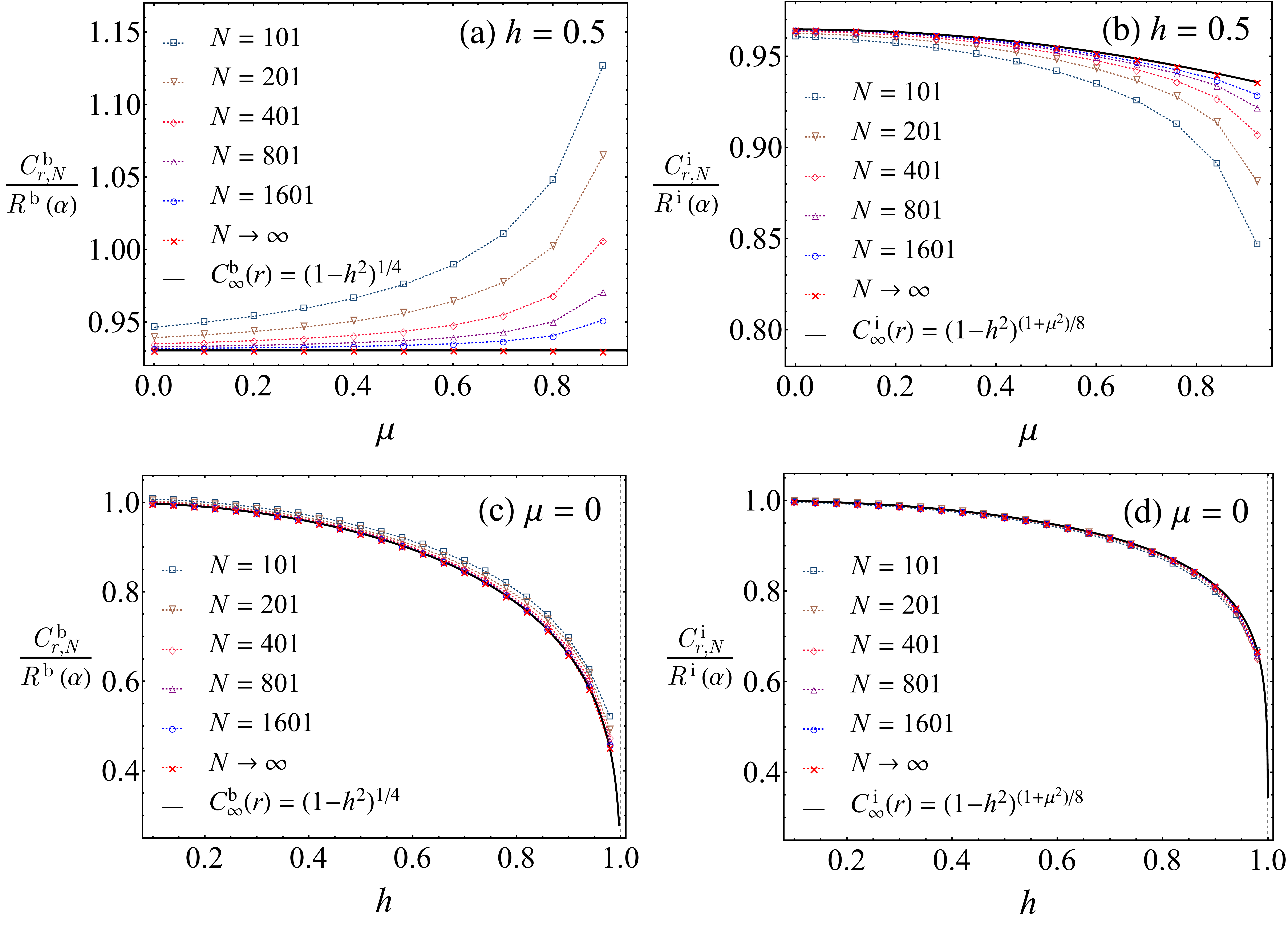}
  \caption{The analysis for extracting the local factors, $C^{\mathrm{b/i}}_{\infty}(r)$, in the non-symmetric TEK phase. The data are sequences for the ratios, $C^{\mathrm{b/i}}_{r,N}/R^{\mathrm{b/i}}(\alpha)$, at a nonlocal distance,$\alpha = r/N \approx 1/8$. The results do not rely on the value of $\alpha$. The selected parameters are: $h=0.5$ in (a) and (b); $\mu=0$ in (c) and (d). This analysis shows that the conjectured approximate expressions in Eqs. (\ref{TEK_Cbr}) and (\ref{TEK_Cir}) are quite good.}
  \label{localfactors}
\end{figure}

There are two types of TEK, one is the symmetric TEK ($\mu=1$), the other is the non-symmetric TEK ($0\leq\mu<1$). In these two TEK, both the bulk and impurity correlations are LRC and can be factorized into a product of \emph{local} and \emph{nonlocal factors}, $C^{\mathrm{b/i}}_{\infty}(r)$ and $R^{\mathrm{b/i}}(\alpha)$, i.e.,
\begin{align}
  C^{\mathrm{b}}(r,\alpha)&= (-1)^{r} C^{\mathrm{b}}_{\infty}(r) R^{\mathrm{b}}(\alpha), \label{TEK_Cb} \\
  C^{\mathrm{i}}(r,\alpha)&= (-1)^{r} C^{\mathrm{i}}_{\infty}(r) R^{\mathrm{i}}(\alpha). \label{}
\end{align}
The local factors are free of $\alpha$, reflecting the local information of the correlations at $\alpha=0$ although $r\gg 1$ as $N\rightarrow\infty$. The nonlocal factors can also be observed when $N$ is finite, i.e.,
\be
  R^{\mathrm{b/i}}(\alpha) \approx R^{\mathrm{b/i}}_{r,N}.
\ee

In the symmetric TEK, there is no impurity at all, i.e. $C^{\mathrm{i}}(r,\alpha)= C^{\mathrm{b}}(r,\alpha)$, previous studies show that the local and nonlocal factors read
\begin{align}
  &C^{\mathrm{b}}_{\infty}(r)=C^{\mathrm{i}}_{\infty}(r)= (1-h^2)^{\frac{1}{4}}, \label{Cinfsymm} \\
  &R^{\mathrm{b}}(\alpha)=R^{\mathrm{i}}(\alpha)= 1-2\alpha. \label{Rb-symmetric}
\end{align}
In fact, the perturbative theory gives the approximate correlation, $C^{\mathrm{b}}(r,\alpha)\approx(-1)^{r} R^{\mathrm{b}}(\alpha)$, which misses the main part of the local factor, $(1-h^2)^{\frac{1}{4}}$. Based on the perturbative theory, the factorizable correlation was conjectured to be true for $0<h<1$ at first \cite{PRE_Vicari_2015}, then proved rigorously \cite{JSM_Li_2016, PRE_Li_2018}, and eventually found to be in deep relation with the FSS analysis \cite{PRE_Li_2019}. Notice that, in Eq. (\ref{Cinfsymm}), the condition $r\gg 1$ is employed, otherwise an exponentially decaying correction, $(1+a~e^{-r/b}/r^{c}) \approx 1$, should be multiplied \cite{JSM_Li_2016}.

We conjecture one might take the same route in the non-symmetric TEK. From the perturbative theory, one can get the nonlocal factors of the bulk and impurity correlations (Appendix \ref{Appendix_B}) \footnote{The factor $R^{\mathrm{b}}(\alpha)$ also appears in a problem with bond defect (impurity) as disclosed in Ref. \cite{PRE_Vicari_2015}, while the factor $R^{\mathrm{i}}(\alpha)$ is a peculiar result for the point impurity in this work.},
\begin{align}
  &R^{\mathrm{b}}(\alpha)= 1-2\alpha-\frac{2}{\pi}\sin(\alpha\pi),  \label{TEK_Rbalpha}\\
  &R^{\mathrm{i}}(\alpha)= 1-2\alpha+\frac{1}{\pi}\sin(2\alpha\pi). \label{TEK_Rialpha}
\end{align}
Then, assuming they are also true if one goes beyond the perturbative theory, one can get the hint to propose appropriate local factors by generalizing the one for the symmetric TEK. By observing the sequences of data on finite lattices, $C^{\mathrm{b/i}}_{r,N}/R^{\mathrm{b/i}}(\alpha)$, we propose the following expressions for the corresponding local factors,
\begin{align}
  &C^{\mathrm{b}}_{\infty}(r)= (1-h^2)^{\frac{1}{4}},  \label{TEK_Cbr}\\
  &C^{\mathrm{i}}_{\infty}(r)= (1-h^2)^{\frac{1+\mu^{2}}{8}}.  \label{TEK_Cir}
\end{align}
Numerical analysis is illustrated in Fig. \ref{localfactors}, which shows that the conjectured expressions are quite good. Thus, we see both the bulk and impurity correlations in TEK are factorizable. Moreover, both of them are LRC, which means that $\xi_{\mathrm{b}}$ and $\xi_{\mathrm{i}}$ are proportional to the system's size.

It seems that the factor $(1-h^2)^{\frac{1}{4}}$ recovers the result of antiferromagnetic phase for the usual situation without ring frustration and its square root gives the famous order parameter, $(1-h^2)^{\frac{1}{8}}$ \cite{sachdev_2011, PR_Yang_1952}, but this is not necessarily true here. In nature, we have demonstrated a phase exhibiting LRC without long-range order (LRO). It is a peculiar phenomenon due to ring frustration. The absence of spontaneous symmetry breaking is ensured by the nondegeneracy of the ground state \cite{JSM_Li_2016}. So the traditional definition of order parameter by the square root of a correlation function is not valid for the TEK. Nonetheless, we can take the whole correlation function as a characteristic at all.

\subsection{KZM-AFZM phase}
\label{KZM-AFZM_phase}
\begin{figure}[t]
  \includegraphics[width=3.4in,angle=0]{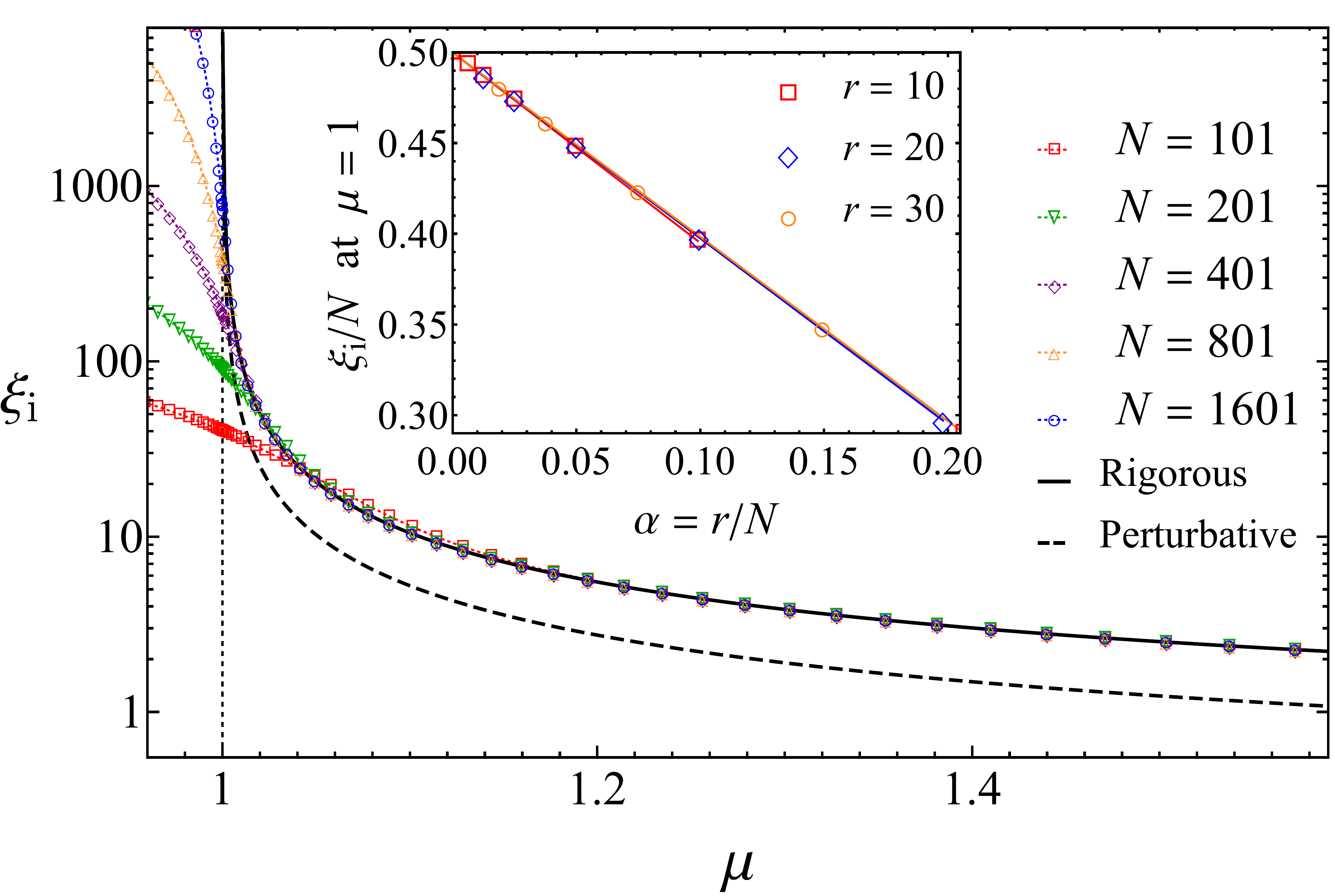}
  \caption{Impurity correlation length $\xi_{\mathrm{i}}$ in the KZM-AFZM phase. The numerical data are obtained by the formula in Eq. (\ref{CL-Numer}) with parameters, $h=0.5$ and varying $\mu$. The solid and dashed lines come from the formula in Eq. (\ref{xi_i})-(\ref{z1_exa}). The inset depicts scaling analysis of $\xi_{\mathrm{i}}$, which shows that we have $\xi_{\mathrm{i}}\approx 0.5 N$ when entering into the TEK at $\mu = 1$. To approach the local distance limit $\alpha \rightarrow 0$, we have fixed $r =10, 20, 30$ so as to extract the result by increasing $N$. In the main plot, only the result for $r=10$ is shown since other cases are similar.}
  \label{xi_KZM}
\end{figure}

In the KZM-AFZM ($\mu>1$ and $h<1$), the correlation functions of the two degenerate KZM states, Eqs. (\ref{KZM-R}) and (\ref{KZM-NS}) (or Eqs. (\ref{approxKZM_R}) and (\ref{approxKZM_NS}) roughly), share the same result, so although the discussion below is based on the state, $|\mathrm{KZM}^{\mathrm{R}}\rangle$, the conclusion is also true for the other state, $|\mathrm{KZM}^{\mathrm{NS}}\rangle$.

In this phase, the bulk correlation is still a LRC and $\xi_{\mathrm{b}}$ is still divergent since Eq. (\ref{TEK_Cb}) holds and
\begin{align}
  &C^{\mathrm{b}}_{\infty}(r)= (1-h^2)^{\frac{1}{4}}, \label{KZM_Cbinf} \\
  &R^{\mathrm{b}}(\alpha) = 1. \label{KZM-Rb}
\end{align}
Nevertheless, the impurity correlation becomes a short-range correlation (SRC),
\be
  C^{\mathrm{i}}(r,\alpha) \approx \left\lbrace
  \begin{array}{llr}
    0, & (\alpha > 0),\\
    g(r)~(-1)^{r} e^{-r/\xi_{\mathrm{i}}}, & (\alpha=0),
  \end{array}\right.
\ee
which means the result is zero if we measure it at a nonlocal distance ($\alpha > 0$) and exponentially decaying if at a local distance ($\alpha = 0$). The prefactor $g(r)$ can be extracted numerically, although it is not easy to find a simple universal expression. The rest part, $(-1)^{r} e^{-r/\xi_{\mathrm{i}}}$, can be captured by the perturbative theory. So we can conclude consistently that we have
\be
  R^{\mathrm{i}}(\alpha) = 0 \label{Requal0}
\ee
in the KZM-AFZM.

The impurity correlation length $\xi_{\mathrm{i}}$ can be extracted by two means. The first is a numerical one. By the formula,
\be  \label{CL-Numer}
  \xi_{\mathrm{i}} = \lim_{N\gg r\gg 1} \left[\ln\left|\frac{C^{\mathrm{i}}_{r,N}}{C^{\mathrm{i}}_{r+1,N}}\right|\right]^{-1},
\ee
we can perform calculations on finite system with $N\gg r\gg 1$ so as to get an extrapolation to infinite $N$. In practical calculations, the limit $\alpha = \frac{r}{N} = 0$ could not be reached exactly. Fortunately, we can fix a small $r$ (say, $r=10, 20, 30$, the results are similar) and increase $N$ (say, $N=101,201,401,801,1601$) so as to figure out the result in the scaling limit $\alpha = \frac{r}{N}\rightarrow 0$. The analysis of the produced numerical data by this means is illustrated in Fig. \ref{xi_KZM}, in which the inset shows the impurity correlation length becomes proportional to the system's size, $\xi_{\mathrm{i}}\approx 0.5 N$, in the local distance limit $\alpha \rightarrow 0$ when entering into the TEK at $\mu = 1$. The second is an analytical one. With the help of perturbative theory (Appendix \ref{Appendix_B}), we can deduce the analytical expression for the impurity correlation length,
\be
  \xi_{\mathrm{i}}=\frac{1}{2|z_{1}|}, \label{xi_i}
\ee
with the approximate solution of the discrete mode,
\be
  z_{1} \approx  i \ln\mu. \label{z1_app}
\ee
But, this solution is good only for small $h$, say, $h\ll 1$. Instead, by substituting the exact solution of $z_{1}$ in Eq. (\ref{z1_exa}) into Eq. (\ref{xi_i}), we found it is excellently coincident with the numerical data for larger $h$ as shown in Fig. \ref{xi_KZM}.

Now we disclose that the $\mathscr{P}_{z}$ symmetry can be broken due to the heavy impurity ($\mu>1$) so that the AFZM forms \footnote{The KZM states induced by point impurity here is different from the ones in a fermionic system as demonstrated in Ref. \cite{PRB_Li_2019}, where spontaneous symmetry breaking won't occur because of the conservation of fermion parity. Please see also Ref. \cite{Kitaev_2008}.}.
By name, AFZM means that the ground state with broken symmetry exhibits antiferromagnetic bulk and localized entangled impurity. We display this phenomenon basing on the definition of two exact AFZM states,
\be
  |\pm\rangle = \frac{1}{\sqrt{2}}\left(|\mathrm{KZM}^{\mathrm{R}}\rangle\pm|\mathrm{KZM}^{\mathrm{NS}}\rangle\right), \label{AFZM}
\ee

Intuitively, the occurrence of spontaneous symmetry breaking can be easily seen in the framework of perturbative theory. By substituting Eqs. (\ref{approxKZM_R}) and (\ref{approxKZM_NS}) into Eq. (\ref{AFZM}), we can clearly see the hierarchical structure of the AFZM by rewriting the two states $|\pm\rangle$ in the form,
\be
  |\pm\rangle \approx \sum_{m=1,2,3,\cdots} \lambda_{m}|\mathrm{bulk},\pm\rangle_{m}\otimes|\mathrm{imp}\rangle_{m},
\ee
where $\lambda_{m}=\sqrt{\mu^{2}-1}/\mu^{m}$, $|\text{bulk},\pm\rangle_{m}$ denote the two antiferromagnetic bulk part of the states,
\be
  &&|\mathrm{bulk},+\rangle_{m} = |\rightarrow_{m},\leftarrow_{m+1},\cdots, \leftarrow_{N-m}\rangle,\\
  &&|\mathrm{bulk},-\rangle_{m} = |\leftarrow_{m},\rightarrow_{m+1},\cdots, \rightarrow_{N-m}\rangle,
\ee
while $|\text{imp}\rangle_{m}$ denote the localized and entangled impurity part that can be hierarchically written down as
\begin{align}
  &|\mathrm{imp}\rangle_{1} = |\rightarrow_{N}\rangle+|\leftarrow_{N}\rangle,\\
  &|\mathrm{imp}\rangle_{2} = |\leftarrow_{N-1},\rightarrow_{N},\leftarrow_{1}\rangle+|\rightarrow_{N-1},\leftarrow_{N},\rightarrow_{1}\rangle,\\
  &|\mathrm{imp}\rangle_{3} = |\rightarrow_{N-2},\leftarrow_{N-1},\rightarrow_{N},\leftarrow_{1},\rightarrow_{2}\rangle+ \nonumber\\
    &~~~~~~~~~~~~~~~~~~~~~~~~|\leftarrow_{N-2},\rightarrow_{N-1},\leftarrow_{N},\rightarrow_{1},\leftarrow_{2}\rangle,\\
  &~~~~~~~~~~\vdots \nonumber
\end{align}
Because $\lambda_{m}$ decreases rapidly with $m$ increasing, the impurity part of the AFZM state is well localized near the impurity. The vast bulk ensures the occurrence of spontaneous symmetry breaking in a spin system \cite{Kitaev_2008}.

\begin{figure}[t]
  \begin{center}
  \includegraphics[width=3.3in,angle=0]{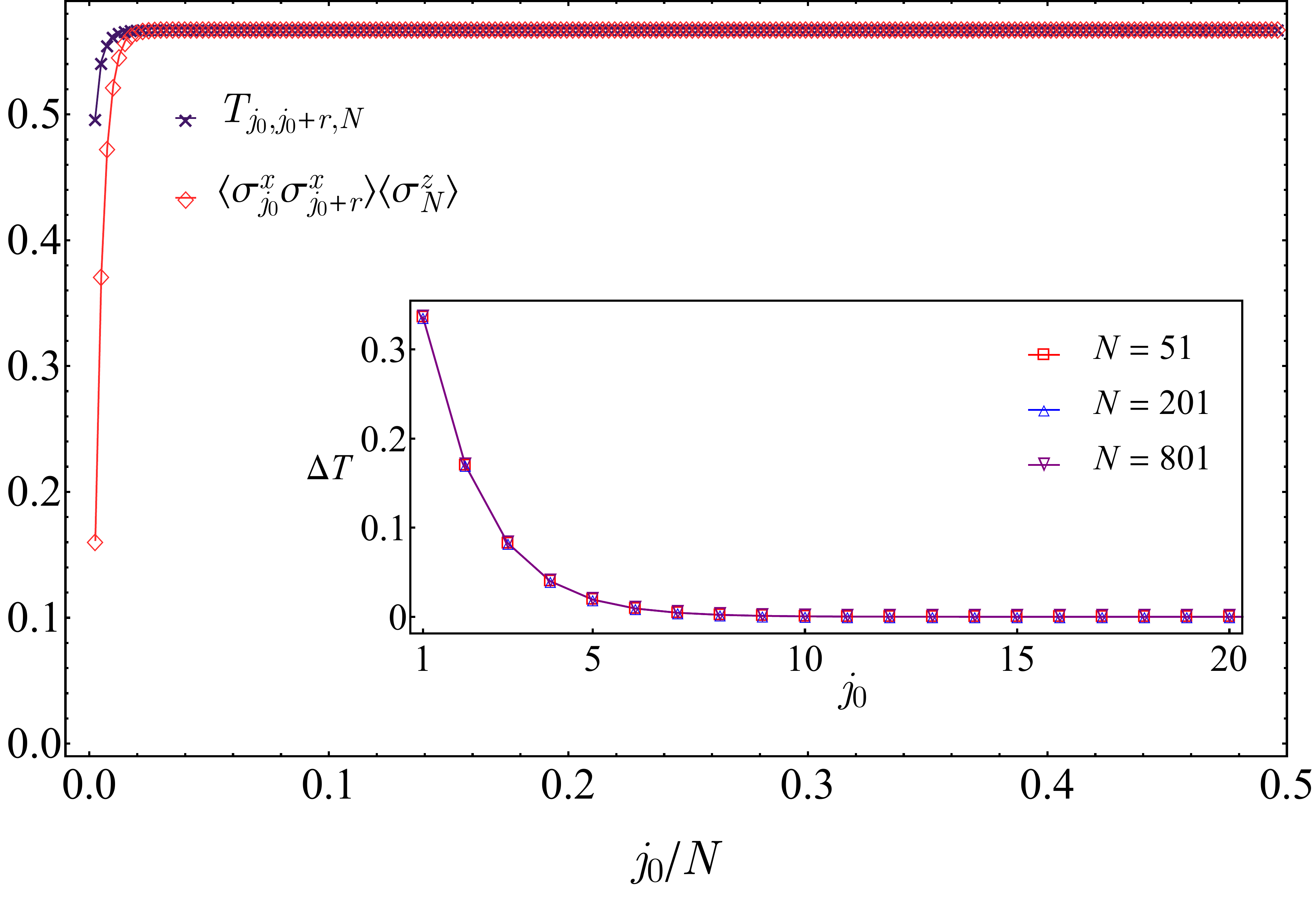}
  \end{center}
  \caption{Three-site correlation function, $T_{j_{0},j_{0}+r,N}$, and the product, $\langle \sigma_{j_{0}}^{x}\sigma_{j_{0}+r}^{x} \rangle\langle \sigma_{N}^{z} \rangle$, for a system with size $N=401$ and parameters, $h=0.1$ and $\mu=1.5$. When $j_{0}$ is far away from the impurity, we have $T_{j_{0},j_{0}+r,N} \approx \langle \sigma_{j_{0}}^{x}\sigma_{j_{0}+r}^{x} \rangle \langle \sigma_{N}^{z} \rangle$, which means that $T_{j_{0},j_{0}+r,N}$ is separable in the thermodynamic limit. However, it is non-separable when site $j_{0}$ is nearby the impurity, which is reflected by the difference, $\Delta T = T_{j_{0},j_{0}+r,N}-\langle \sigma_{j_{0}}^{x}\sigma_{j_{0}+r}^{x} \rangle\langle \sigma_{N}^{z} \rangle$, as shown in the inset. We see that $\Delta T$ is insensitive to the system's size nearby the impurity, which reflects the stable and non-separable localized part of the ground state.}
  \label{three_site}
\end{figure}

Inspired by this perturbative picture, we investigate rigorously the three-site correlation, $T_{j_{0},j_{0}+r,N}=\langle \sigma_{j_{0}}^{x}\sigma_{j_{0}+r}^{x} \sigma_{N}^{z} \rangle$, in which $j_{0}$ is defined in Eq. (\ref{j0}). The numerical result is illustrated in Fig. \ref{three_site}. We see that $T_{j_{0},j_{0}+r,N}$ can be separated into a product of two-site correlation and one-site average,
\begin{align}
  \langle \sigma_{j_{0}}^{x}\sigma_{j_{0}+r}^{x} \sigma_{N}^{z} \rangle \approx \langle \sigma_{j_{0}}^{x}\sigma_{j_{0}+r}^{x} \rangle \langle \sigma_{N}^{z} \rangle ,
\end{align}
when both $N$ and $j_{0}$ are large enough ($j_{0}/N \neq 0$ as $N\rightarrow\infty$, i.e. site $j_{0}$ is far away from the impurity). This separability  signifies a possible symmetry breaking in the vast bulk, because the two-site correlation can be further separated in the AFZM states,
\be
  \langle\pm |\sigma_{j_{0}}^{x}\sigma_{j_{0}+r}^{x}|\pm \rangle \approx \langle \pm| \sigma_{j_{0}}^{x} | \pm\rangle\langle\pm|\sigma_{j_{0}+r}^{x}|\pm \rangle.
\ee
Thus in the bulk part of the AFZM states, the $\mathscr{P}_{z}$ symmetry is broken and we can introduce the order parameter in the usual way \cite{PR_Yang_1952},
\be
  m_{x} = \sqrt{|\langle \pm|\sigma_{j_{0}}^{x}\sigma_{j_{0}+r}^{x}|\pm \rangle|}=(1-h^{2})^\frac{1}{8},
\ee
since the rigorous two-site correlation is given by Eq. (\ref{KZM_Cbinf}). Nevertheless, $T_{j_{0},j_{0}+r,N}$ is still non-separable for small $j_{0}$ ($j_{0}/N\rightarrow 0$ as $N\rightarrow\infty$, i.e. site $j_{0}$ is nearby the impurity) as shown in the inset of Fig. \ref{three_site}. This result is in good agreement with the one by perturbative theory, which means that the AFZM states remain entangled nearby the impurity.

\subsection{Transition from TEK to KZM-AFZM}

\begin{figure}[t]
  \begin{center}
	\includegraphics[width=3.4in,angle=0]{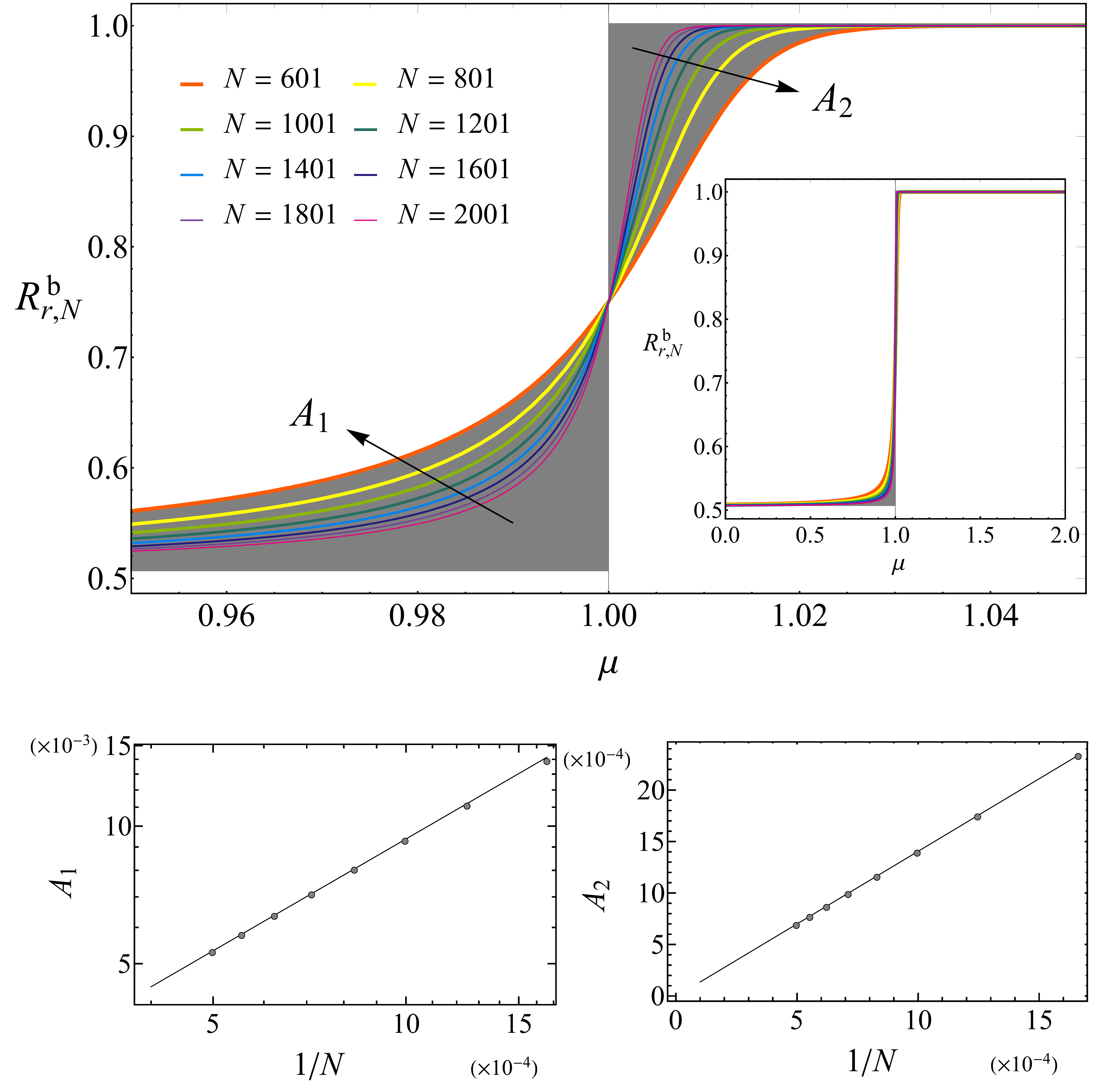}
  \end{center}
  \caption{Upper plot: FSS analysis of the nonlocal factor of the bulk correlation function in the transition from TEK to KZM-AFZM phases, which shows that the sequence of data, $R^{\mathrm{b}}_{r,N}$, approaches the steplike function in Eq. (\ref{step-like-Rb}), $R^{\mathrm{b}}(\alpha)$, with $N$ increasing. The shaded areas, $A_{1}$ and $A_{2}$, defined in Eqs. (\ref{A1}) and (\ref{A2}) indicate the difference between $R^{\mathrm{b}}_{r,N}$ and $R^{\mathrm{b}}(\alpha)$. The parameters, $h=0.5$ and $\alpha=1/8$, are selected in the plot. Inset: the same plot in a wider scope of the parameter, $\mu\in(0,2)$. Lower left plot: FSS analysis for $A_{1}$. Lower right plot: FSS analysis for $A_{2}$.   }
  \label{step}%
\end{figure}

According to the discussion above, the features of the transition from TEK to KZM-AFZM can be reflected in both the bulk and impurity correlations.

First, the bulk correlation remains to be LRC and its local factor does not change with varying $\mu$ according to Eqs. (\ref{TEK_Cbr}) and (\ref{KZM_Cbinf}), nevertheless, its nonlocal factor is a steplike function according to Eqs. (\ref{Rb-symmetric}), (\ref{TEK_Rbalpha}), and (\ref{KZM-Rb}). For convenience, we write down it explicitly,
\be
  &R^{\mathrm{b}}(\alpha)=\left\lbrace
  \begin{array}{llr}
    1-2\alpha-\frac{2}{\pi}\sin(\alpha\pi), & (\mu<1),\\
    1-2\alpha, & (\mu=\mu_{\mathrm{c}}=1),\\
    1, & (\mu>1).
  \end{array}\right. \label{step-like-Rb}
\ee
In fact, this is a direct result by taking $N\rightarrow\infty$ in the perturbative theory (Appendix \ref{Appendix_B}). It is interesting to verify how this steplike function is approached by the rigorous data of finite-size systems in a consistent way. We can numerically extract the finite-size version of the nonlocal factor by $R^{\mathrm{b}}_{r,N}=C^{\mathrm{b}}_{r,N}/(1-h^2)^{\frac{1}{4}}$ so as to carry out the FSS analysis. In Fig. \ref{step}, we label the shaded areas, $A_{1}$ and $A_{2}$, defined by
\be
  A_{1} = \int_{0}^{1} |R^{\mathrm{b}}_{r,N}-R^{\mathrm{b}}(\alpha)| \mathrm{d}\mu, \label{A1} \\
  A_{2} = \int_{1}^{\infty} |R^{\mathrm{b}}_{r,N}-R^{\mathrm{b}}(\alpha)| \mathrm{d}\mu. \label{A2}
\ee
The two insets in Fig. \ref{step} show the scaling behavior,
\be
  A_{1}&\approx& 1.41 N^{-1},\\
  A_{2}&\approx& 2.57 N^{-0.81}.
\ee

Second, the impurity correlation alters from LRC to SRC. According to Eqs. (\ref{Rb-symmetric}), (\ref{TEK_Rialpha}), and (\ref{Requal0}), the nonlocal factor is likewisely a steplike function and reads
\be
  &R^{\mathrm{i}}(\alpha)=\left\lbrace
  \begin{array}{llr}
    1-2\alpha+\frac{2}{\pi}\sin(\alpha\pi), & (\mu<1),\\
    1-2\alpha, & (\mu=\mu_{c}=1),\\
    0, & (\mu>1).
  \end{array}\right. \label{step-like-Ri}
\ee
However, as a better characterization, we can observe that the impurity correlation length $\xi_{\mathrm{i}}$ undergoes a transition from a divergent value to a finite one as illustrated in Fig. \ref{xi_KZM}.

\subsection{Difference between PM-1 and PM-2 subphases}

In the PM-1 and PM-2 subphases, both bulk and impurity correlation functions are SRC with finite correlation length. It is hard to discern the difference of the two subphases through the correlation function of the ground state, because the behaviour of the correlation lengths are found to be the same,
\be
  \xi_{\mathrm{b}} \text{  or  } \xi_{\mathrm{i}} \sim \frac{1}{\ln h}, \label{ln_h}
\ee
in both PM-1 and PM-2. We have confirmed this behaviour by rigorous calculation on lattices with large enough $N$. Whereas, the difference can be easily discerned in Fig. \ref{schematics}(d), which shows that the first excited state in PM-1 is a discrete energy level, while in PM-2 it becomes the bottom of the continuous band. Correspondingly, we can resort to the first excited state to distinguish the two subphases.

For simplicity, let us see a perturbative theory that is fit for the PM phases. In PM-1 ($\mu<1$), the ground state and the first excited state evolve adiabatically into the following simple states respectively,
\be
  |E_{0}^{\mathrm{R}}\rangle &\rightarrow& |N_{\uparrow}\rangle =|\uparrow_{1},\cdots,\uparrow_{N-1},\uparrow_{N}\rangle, \label{one_spin_up_N}\\
  |E_{1'}^{\mathrm{NS}}\rangle &\rightarrow& |N_{\downarrow}\rangle = |\uparrow_{1},\cdots,\uparrow_{N-1},\downarrow_{N}\rangle. \label{one_spin_down_N}
\ee
in the limit $h\rightarrow\infty$. Obviously, the ground state $|E_{0}^{\mathrm{R}}\rangle$ behaves extended, while the first excited state $|E_{1'}^{\mathrm{NS}}\rangle$ shows a localization behaviour around the impurity at site $N$. When $\mu\neq 0$ and $h$ is finite, we can do the perturbative calculations in subspaces with appropriate parity. Although the ground state will be blended with many states of odd parity, there is no chance to develop a localized mode. While for the first excited state, we can perform perturbative calculation in the one-spin-down subspace with even parity $(n=1,2,\cdots,N)$,
\be
  \{|n_{\downarrow}\rangle=|\cdots,\uparrow_{n-1},\downarrow_{n},\uparrow_{n+1},\cdots\rangle\}.
\ee
Because the state in Eq. (\ref{one_spin_down_N}) dominates for $0<\mu<1$, we get a localized state,
\be
  |E_{1'}^{\mathrm{NS}}\rangle = \sum_{n=1}^{N} c_{n} |n_{\downarrow}\rangle,
\ee
where
\be
&c_{n}=A \times\left\lbrace
\begin{array}{llr}
	\frac{(-1)^{n}h^{-n}}{(2-2\mu)^{n}}, & (1\leq n\leq\frac{N-1}{2}),\\
	\frac{-(-1)^{n}h^{n-N}}{(2-2\mu)^{N-n}}, & (\frac{N+1}{2}\leq n<N),\\
	1         ,                 & (n=N ),
\end{array}\right.
\ee
with $A=\sqrt{\frac{4 h^2(1-\mu)^{2}-1}{ 4 h^2(1-\mu)^{2}+1}}$. Consistently, it is easy to recover Eq. (\ref{one_spin_down_N}) by setting $h\rightarrow\infty$. However, in PM-2 ($\mu>1$), the first excited state changes non-adiabatically due to energy level crossing, $|E_{1'}^{\mathrm{NS}}\rangle\rightarrow|E_{1}^{\mathrm{NS}}\rangle$, so we get an extended state instead,
\be
  |E_{1}^{\mathrm{NS}}\rangle = \sqrt{\frac{2}{N}}\sum_{n=1}^{N} (-1)^{n-1} \sin\frac{n\pi}{N} |n_{\downarrow}\rangle.
\ee

\subsection{Transitions from TEK to PM-1 and KZM-AFZM to PM-2}

\begin{figure}[t]
  \begin{center}
	\includegraphics[width=3.4in,angle=0]{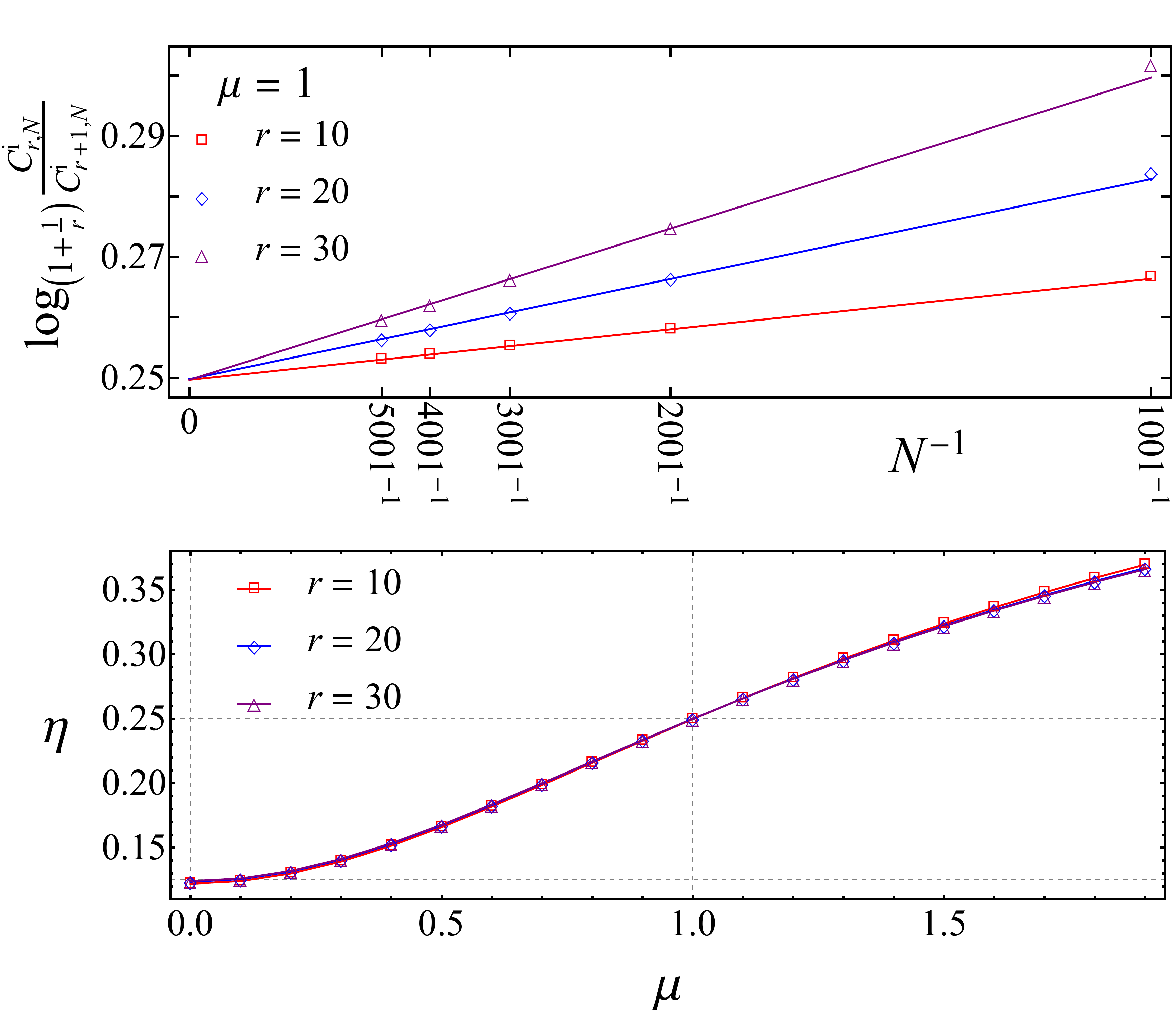}
  \end{center}
  \caption{Upper plot: FSS analysis of the value of $\eta$ at $\mu=1$ and $h=1$ basing on the formula in Eq. (\ref{eta}). The data are produced on systems with $N$ ranging from 1000 to 5000. By extrapolating to the limit $N\rightarrow\infty$, we obtain $\eta \approx 0.249685, 0.249803, 0.249693$ for fixed $r =10, 20, 30$ respectively. So we see that the well-known exact result $\eta=1/4$ is reproduced very accurately. Lower plot: the values of $\eta$ versus $\mu$ along the critical line $h=1$ (Please see the phase diagram in Fig. \ref{lowlevels}), which is obtained by using the formula in Eq. (\ref{eta}) in the same way. The grid lines show that we obtain $\eta \approx 1/4 $ at $\mu=1$ and  $\eta \approx 1/8$ at $\mu=0$.}
  \label{fig_eta}
\end{figure}

In the transition from the gapless TEK to gapped PM-1, both the bulk and impurity correlations change from LRC to SRC. The ground state and the first excited state evolve adiabatically (Please see Fig. \ref{schematics}(c)). Two gaps open at the same time when entering into the PM-1,
\be
  \Delta_{2} &=& 2 \sqrt{1+h^{2}-2h\cosh |z_{3}|},\\
  \Delta_{1} &=& 2(h-1) -\Delta_{2},
\ee
where
\be
  z_{3} = i \ln\frac{h(1-\mu^{2})+\sqrt{4\mu^{2}+h^{2}(1-\mu^{2})^{2}}}{2} \label{z3_exa}
\ee
is a complex root in the even channel $P_{z}^{+} H^{\mathrm{NS}}$ as described in Eq. (\ref{z3}). Notice that we get $\Delta_{1}=\Delta_{2}=0$ at $h=1$.

While in the transition from the gapped KZM-AFZM to gapped PM-2, the ground state and the first excited state also evolve adiabatically (Fig. \ref{schematics}(b)). The bulk correlation changes from LRC to SRC, while the impurity correlation keeps SRC. The two gaps close and open in a reverse direction,
\be
  &\Delta_{1} = 2|h-1| \Theta(1-h), \\
  &\Delta_{2} = 2|h-1| \Theta(h-1),
\ee
where $\Theta(x)$ is the Heaviside step function. The impurity correlation length $\xi_{\mathrm{i}}$ behaves differently in the two phases according to Eqs. (\ref{xi_i}) and (\ref{ln_h}).

The above two transitions occur on the same critical line, $h=1$, on which the two gaps close at the same time. It turns out that the local factor of the impurity correlation exhibits a power law,
\be
  C_{\infty}^{\mathrm{i}}(r) \approx \frac{b}{r^{\eta}},
\ee
along the whole transition line. $b$ and $\eta$ are constants relying on $\mu$.
To extract the value of $\eta$, we carry out FSS analysis basing on the formula,
\be
  \eta = \lim_{N\gg r\gg 1}\log_{(1+\frac{1}{r})}\frac{C_{r,N}^{\mathrm{i}}}{C_{r+1,N}^{\mathrm{i}}}. \label{eta}
\ee
The merit of this method is that the nonlocal factor in $C_{r,N}^{\mathrm{i}}$ is cancelled as long as $N\gg r\gg 1$ is fulfilled. First, as a test, we work on the system with $\mu=1$ and $h=1$, where an exact result, $\eta=1/4$, is well-known \cite{PRE_Li_2019}. In the upper plot of Fig. \ref{fig_eta}, we show that this method can reproduce the exact value very accurately. Because the impurity correlation decays very slow due to the power law, we can perform steady calculations on systems as large as $N=5000$. Then, using this method, we work out the values of $\eta$ along the critical line by varying $\mu$, which are illustrated in the lower plot of Fig. \ref{fig_eta}.

At last, we address the classical impurity limit ($\mu=0$) \cite{Annals_Nersesyan_2016}, where interesting physics due to ring frustration is quite different from that of the OBC case. In this limit, the ground state $|E_{0}^{R}\rangle$ and the first excited state $|E_{1'}^{NS}\rangle$ still evolve adiabatically during the transition, but they become absolutely degenerate because $\sigma_{N}^{x}$ commutes with the system Hamiltonian, $[\sigma_{N}^{x},H] = 0$. $|E_{0}^{R}\rangle$ and $|E_{1'}^{NS}\rangle$ exhibit good quantum number of $\mathscr{P}_{z}$. They can be mixed so as to become the eigenstates of $\sigma_{N}^{x}$ (as well as the parity in $x$ direction, $\mathscr{P}_{x} = \prod_{j=1}^{N}(-\sigma_{j}^{x})$),
\be
  \sigma_{N}^{x}(|E_{0}^{R}\rangle \pm |E_{1'}^{NS}\rangle) = \pm(|E_{0}^{R}\rangle \pm |E_{1'}^{NS}\rangle).
\ee
But to superpose the two states is just a manipulation on the quantum states, which does not mean any phenomenon of spontaneous symmetry breaking along the whole line $\mu=0$, because the resulting states are entangled. As can be roughly seen by Eqs. (\ref{one_spin_up_N}) and (\ref{one_spin_down_N}). Thus, this situation is in contrast with the OBC case \cite{PRB_Plastina_2017}, or the AFZM discussed in Sec. \ref{KZM-AFZM_phase}.

\section{Summary}

In summary, we have studied the phases and transitions in the quantum Ising ring as a result of the interplay between point impurity and ring frustration. By rigorous solution and complimentary perturbative theory, we have constructed the ground-state phase diagram. The phases and transitions are characterized by appropriate bulk and impurity correlation functions in the context of nonlocality in the thermodynamic limit. Specifically, the two-site correlation function can be factorized into a product of local and nonlocal factors if it is a LRC. In gapless TEK phase, the spontaneous symmetry breaking is absent because the ground state is unique, which indicates that the LRC does not lead to LRO in this peculiar situation. Whereas, the spontaneous symmetry breaking does occur in the gapped KZM-AFZM phase, where the doubly degenerate KZM states can transform into two AFZM states. A type of three-site correlation function can be utilized to capture this phenomenon. In a AFZM state, LRO develops in the vast bulk and the entangled part of the state is retained locally near the impurity. In the transition from TEK to KZM-AFZM, both the bulk and impurity nonlocal factors of the correlations are found to be steplike functions in the thermodynamic limit. In contrast, the impurity correlation changes from LRC to SRC by going through a power law critical line for the local factor in the transition from TEK to PM-1.

\section*{ACKNOWLEDGMENTS}
This work is supported by NSFC under Grants No. 11074177.

\appendix

\section{Diagonalization of $H_{\mathrm{R/NS}}$} \label{Appendix_A}

The free fermion Hamiltonians with PBC/APBC are
\be \label{app-2-17}
  H^{\mathrm{R/NS}}=\sum_{i,j}\left[ f_{i}^{\dagger}A_{ij} f_{j}+\frac{1}{2}(f_{i}^{\dagger}B_{ij}f^{\dagger}_{j}+h.c.) \right].
\ee
We have omitted the superscript R/NS for abbreviation, $A=A^{\mathrm{R/NS}}$ and $B=B^{\mathrm{R/NS}}$. Matrix $A$ is Hermitian, while the matrix $B$ is antisymmetric. At the same time, both $A$ and $B$ are real. We try to find a linear transformation,
\begin{equation} \begin{aligned}
\eta_{q}&=\sum_{j}(g_{q,j}f_{j}+h_{q,j}f_{j}^{\dagger}), \\
\eta_{q}^{\dagger}&=\sum_{j}(g_{q,j}f_{j}^{\dagger}+h_{q,j}f_{j}),
\end{aligned}
\end{equation}
with the canonical coefficients that will lead to the diagonalized form for the Hamiltonians,
\begin{eqnarray}
H^{\mathrm{R/NS}}=\sum_{q}\omega_{q}\eta_{q}^{\dagger}\eta_{q} + \mathrm{constant},
\end{eqnarray}
We introduce two matrices, $\Phi$ with elements $\phi_{q,j}=g_{q,j}+h_{q,j}$ and $\Psi$ with elements $\psi_{q,j}=g_{q,j}-h_{q,j}$.
This leads to the eigenvalue problem,
\begin{equation*}
\begin{aligned}
\omega^{2}\Phi=&\Phi(A-B)(A+B)\\
\omega^{2}\Psi=&\Psi(A+B)(A-B)
\end{aligned}
\end{equation*}
where $\omega=\mathrm{diag}(\omega_{1},\omega_{2},\cdots,\omega_{N})$ is a diagonal matrix that consists of eigenvalues.
Lengthy, but straightforward, calculations lead to the equation in Eq. (\ref{Lieb_eqn}), which gives the solution of the roots of $q$ and the spectrum, $\omega(q)=2\sqrt{ 1+h^{2}-2h \cos q}$. Most roots of $q$ are real, except for the ones denoted by $z=z_{1/2/3/4}$ as described in the main text. They may be complex and lead to discrete modes as listed in Table \ref{table_1}, where $h_{\mu}=e^{\sgn(1-\mu)|z_{1}|}$ denotes a line fulfilling $\omega(z_1)=2\sqrt{1+h_{\mu}^{2}-2h_{\mu} \cos z_1}=0$.
\begin{table}[!htbp]
	\renewcommand{\arraystretch}{1.5} 
	\centering
	\begin{tabular}{p{3.5cm}<{\centering}p{1.cm}<{\centering}p{1.cm}<{\centering}p{1.cm}<{\centering}}
		\hline
		\hline
		$\Omega_{z_{1}}$&$\Omega_{z_{2}}$&$\Omega_{z_{3}}$&$\Omega_{z_{4}}$\\
		\hline
		$\sgn(h_{\mu}-h)\omega(z_{1})$&$\omega(z_{2})$&$\omega(z_{3})$&$-\omega(z_{4})$\\
		\hline
	\end{tabular}
	\caption{Four possible discrete modes. }
	\label{table_1}
\end{table}

The elements of $\Phi$ and $\Psi$ can be worked out as
\begin{align}
\phi_{q,j}&=c_{1}\left[ \sin qj-\mu\mathscr{P}_{z}\sin(N-j)q\right], \\
\psi_{q,j}&=c_{2}\frac{\mu\sin qj-\mathscr{P}_{z}\sin(N-j)q}{1+(\mu-1)\delta_{j,N}},
\end{align}
where $c_{1/2}$ are normalization factors.

\section{Perturbative theory for the TEK and KZM-AFZM} \label{Appendix_B}

The perturbative theory is carried out in the subspace composed of the one-kink Ising states in Eqs. (\ref{jright}) and (\ref{jleft}), which are also eigenstates of the main Hamiltonian $H_{0}=\sum_{j=1}^{N}\sigma_{j}^{x}\sigma_{j+1}^{x}$. As a perturbation ($h\ll 1$), the transverse term $V=H-H_{0}$ does not commute with $H_{0}$, so we can get approximate eigenstates by diagonalizing it in this subspace. Because of the commutator $[\mathscr{P}_{z},H]=0$, we introduce another set of states with good quantum number of $\mathscr{P}_{z}$,
\be
  &&|j,\uparrow \rangle= \frac{1}{\sqrt{2}}\left(|j,\rightarrow\rangle+|j,\leftarrow\rangle \right),\\
  &&|j,\downarrow \rangle= \frac{1}{\sqrt{2}}\left(|j,\rightarrow\rangle-|j,\leftarrow\rangle\right),
\ee
so that the effective Hamiltonian is obtained as
\be
  H_{\mathrm{eff}}=\sum_{j=1}^{N}[(2-N)(|j,\uparrow\rangle\langle j,\uparrow|+|j,\downarrow\rangle\langle j,\downarrow|) \nonumber\\
  - h_{j+1}(|j,\uparrow\rangle\langle j+1,\uparrow|-|j,\downarrow\rangle\langle j+1,\downarrow|+\mathrm{h.c.})]
\ee
where $h_{j}=h+(\mu-1)h\delta_{j,N}$.
The ground state can be worked out as
\begin{eqnarray}
  |E_{0}^{\mathrm{R}}\rangle \approx \sum_{j=1}^{N-2}  \left[ \cos k_{0}j +\cos(N-j-1)k_{0} \right]|j,\uparrow \rangle \nonumber\\
  + \sum_{j=N-1}^{N}\frac{\left[ \cos k_{0} +\cos(N-2)k_{0} \right] }{2\cos k_{0}-\mu}|j,\uparrow \rangle, \label{E_0_approx}
\end{eqnarray}
where
\be
  k_{0}=\left\lbrace
  \begin{array}{lll}
    &\frac{\pi(1-\mu)}{N(1-\mu)+(1+\mu)}\approx\frac{\pi}{N},& (\mu<1),\\
    &\frac{1}{N}\arccos(\frac{2\mu-N(1-\mu^2)}{1+\mu^{2}}), & (\mu\approx 1),\\
    &i\ln\mu, & (\mu>1).
  \end{array}   \right. \label{k0}
\ee
In the KZM-AFZM, the ground state is doubly degenerate. The above ground state plays the role of a KZM state with odd parity, $|\mathrm{KZM}^{\mathrm{R}}\rangle=|E_{0}^{\mathrm{R}}\rangle$. Another KZM state with even parity is worked out as
\be
  |\mathrm{KZM}^{\mathrm{NS}}\rangle\approx\sum_{j=1}^{N-2} \left[ \sin k_{0}j -\sin(N-j-1)k_{0} \right)|j,\downarrow \rangle \nonumber\\
  + \sum_{j=N-1}^{N}(-1)^j\frac{\left( \sin k_{0}- \sin(N-2)k_{0} \right] }{2\cos k_{0}+\mu}|j,\downarrow \rangle.
\ee

In the thermodynamic limit $N\rightarrow\infty$,  the two degenerate KZM states can be simplified as
\be
  |\mathrm{KZM}^{\mathrm{R}}\rangle=\sum_{j=1}^{N}\psi_{j}|j, \uparrow\rangle,
\ee
\be
  |\mathrm{KZM}^{\mathrm{NS}}\rangle=\sum_{j=1}^{N}\chi_{j}|j, \downarrow \rangle,
\ee
where the coefficients read
\be
  &\psi_{j}=\frac{\sqrt{\mu^2-1}}{2\mu}\times\left\lbrace
  \begin{array}{llr}
    \mu^{-j}, & (1\leq j \leq \frac{N-1}{2}),\\
    \mu^{-N+j+1}, & (\text{other } j),\\
    1, & (j=N-1,N),
  \end{array}\right.
\ee
\be
  &\chi_{j}=\left\lbrace
  \begin{array}{llr}
    (-1)^{j}\psi_{j}, & (1\leq j \leq \frac{N-3}{2}),\\
    (-1)^{j-1}\psi_{j}, & (\text{other } j),\\
    0, & (j = \frac{N-1}{2}),
  \end{array}\right.
\ee

The longitudinal correlation function of the ground state defined in Eq. (\ref{Cxx}) can also be worked out in the framework of perturbative theory. By Eq. (\ref{E_0_approx}), we get
\be
  C_{j,j+r} \approx (-1)^{r}  F(k_{0}) G(k_{0}), \label{Capp}
\ee
in which

\begin{widetext}
\be
  &&F(k_{0})=\frac{2\cos^{2}(\frac{N- 1}{2}k_{0})}{\frac{2}{N}\left[\frac{\cos k_{0}+\cos(N-2) k_{0}}{2\cos k_{0}-\mu} \right]^{2}+\frac{1}{N}\sum_{l=1}^{N-2}\left[\cos l k_{0}+\cos(N-l-1)k_{0}\right]^{2}}\\
  &&G(k_{0})=1-2 \frac{r}{N}+ \frac{\sin(N-2r-2j)k_{0}+\sin N k_{0}-\sin(N-2j)k_{0}}{N\sin k_{0}}
\ee
Substituting $k_0$ in Eq. ({\ref{k0}}) into $F(k_{0})$ and $G(k_{0})$ and taking $N\rightarrow\infty$, we find that
\be
  F(k_{0})\rightarrow\left\lbrace
  \begin{array}{lll}
    &1,& (\mu<1),\\
    &\frac{1}{2}, & (\mu=1),\\
    &\frac{N(\mu^{2}-1)}{2\mu e^{N\ln\mu}}, & (\mu>1),
  \end{array}   \right.
  \label{Fk0}
\ee
\be
  G(k_{0})\rightarrow\left\lbrace
  \begin{array}{lll}
    &(1-2\alpha)+\frac{\sin[(1-2\alpha-2\frac{j}{N})\pi]-\sin[(1-2\frac{j}{N})\pi]}{\pi},& (\mu<1),\\
    &2(1-2\alpha), & (\mu=1),\\
    &1-2 \frac{r}{N}+ \frac{2\mu e^{N\ln\mu}}{N(\mu^{2}-1)} [1-e^{-2(N-r-j)\ln\mu}+e^{-2(r+j)\ln\mu}+e^{-2(N-j)\ln\mu}-e^{-2j\ln\mu}], & (\mu>1).
  \end{array}   \right.
  \label{Gk0}
\ee
By substituting Eqs. (\ref{Fk0}) and (\ref{Gk0}) into (\ref{Capp}), we can recover the nonlocal factors for the long-range correlations in the TEK ($\mu<1$) and the exponentially decaying short-range impurity correlation in the KZM-AFZM ($\mu>1$) as described in the main text.

\section{Determinant representation of correlation functions} \label{Appendix_C}

By using Wick's theorem and Majorana fermions,
\begin{align}
  A_{j}=f^{\dagger}_{j}+f_{j},\quad B_{j}=f^{\dagger}_{j}-f_{j},
\end{align}
the two-site longitudinal correlation function defined in Eq. (\ref{Cxx}) can be represented by a $r$-th order determinant. First, we can write down,
\begin{equation}
\begin{aligned}
C_{j,j+r}=&\langle 0^{\mathrm{R}} |\eta_{z_{1}} B_{j}A_{j+1}B_{j+1}A_{j+2}\cdots A_{j+r-1}B_{j+r-1}A_{j+r}\eta_{z_{1}}^{\dagger}| 0^{\mathrm{R}} \rangle\\
=&\langle\eta_{z_{1}}\eta_{z_{1}}^{\dagger}\rangle\langle B_{j}A_{j+1}B_{j+1}A_{j+2}\cdots A_{j+r-1}B_{j+r-1}A_{j+r}\rangle\\
&+\left(\langle\eta_{z_{1}}B_{j}\rangle\langle A_{j+1}\eta_{z_{1}}^{\dagger}\rangle-\langle\eta_{z_{1}}A_{j+1}\rangle\langle B_{j}\eta_{z_{1}}^{\dagger}\rangle \right) \langle B_{j+1}A_{j+2}B_{j+2}\cdots B_{j+r-1}A_{j+r}\rangle\\
&+\left(\langle\eta_{z_{1}}B_{j}\rangle\langle A_{j+2}\eta_{z_{1}}^{\dagger}\rangle-\langle\eta_{z_{1}}A_{j+2}\rangle\langle B_{j}\eta_{z_{1}}^{\dagger}\rangle \right) \langle A_{j+1}B_{j+2}A_{j+2}B_{j+2}\cdots B_{j+r-1}A_{j+r}\rangle\\
&+\cdots\cdots.
\end{aligned}
\end{equation}
Then by the contractions,
\begin{equation}
\begin{aligned}
&\langle\eta_{z_{1}}\eta_{z_{1}}^{\dagger}\rangle=1,\\
&\langle A_{i}A_{j}\rangle=-\langle B_{i}B_{j}\rangle=\delta_{i,j}, \\
&\langle B_{j}A_{j+r}\rangle\equiv G_{j,j+r}=\sum_{q}(h_{q,j}-g_{q,j})(h_{q,j+r}+g_{q,j+r}),  \\
&\langle\eta_{z_{1}}A_{j+r}\rangle\langle B_{j}\eta_{z_{1}}^{\dagger}\rangle-\langle\eta_{z_{1}}B_{j}\rangle\langle A_{j+r}\eta_{z_{1}}^{\dagger}\rangle\equiv F_{j,j+r}=2(h_{z_{1},j}-g_{z_{1},j})(h_{z_{1},j+r}+g_{z_{1},j+r}),\\
\end{aligned}
\end{equation}
we can arrive at
\begin{equation}
	C_{j,j+r}=\det\left[  \begin{array}{ccccc}
	G_{j,j+1}-F_{j,j+1}&G_{j,j+2}-F_{j,j+2}&\cdots\cdots&G_{j,j+r}-F_{j,j+r}\\
	G_{j+1,j+1}-F_{j+1,j+1}&G_{j+1,j+2}-F_{j+1,j+2}&\cdots\cdots&G_{j+1,j+r}-F_{j+1,j+r}\\
	&\cdots\cdots&&\\
    G_{j+r-1,j+1}-F_{j+r-1,j+1}&G_{j+r-1,j+2}-F_{j+r-1,j+2}&\cdots\cdots&G_{j+r-1,j+r}-F_{j+r-1,j+r}\\
	\end{array}\right] . \label{app_Cxx}
\end{equation}
Except for the translational symmetric case ($\mu=1$), this determinant is generally not a Toeplitz determinant due to the presence of impurity ($\mu\neq 1$), however, it can still be evaluated numerically for quite large systems.

The three-site correlation function defined in Eq. (\ref{Tzxx}) can also be represented by a determinant. By Wick's theorem, we have
\begin{equation}
\begin{aligned}
T_{j,j+r,N}=&-\langle 0^{\mathrm{R}} |\eta_{z_{1}}B_{N}A_{N} B_{j}A_{j+1}B_{j+1}A_{j+2}\cdots A_{j+r-1}B_{j+r-1}A_{j+r}\eta_{z_{1}}^{\dagger}| 0^{\mathrm{R}} \rangle\\
=&-\langle\eta_{z_{1}}\eta_{z_{1}}^{\dagger}\rangle\langle B_{N}A_{N}B_{j}A_{j+1}B_{j+1}A_{j+2}\cdots A_{j+r-1}B_{j+r-1}A_{j+r}\rangle\\
&-\left(\langle\eta_{z_{1}}B_{N}\rangle\langle A_{N}\eta_{z_{1}}^{\dagger}\rangle-\langle\eta_{z_{1}}A_{N}\rangle\langle B_{N}\eta_{z_{1}}^{\dagger}\rangle\right) \langle B_{j}A_{j+1}B_{j+1}\cdots B_{j+r-1}A_{j+r}\rangle\\
&-\left(\langle\eta_{z_{1}}B_{N}\rangle\langle A_{j+1}\eta_{z_{1}}^{\dagger}\rangle-\langle\eta_{z_{1}}A_{j+1}\rangle\langle B_{N}\eta_{z_{1}}^{\dagger}\rangle\right) \langle A_{N}B_{j}B_{j+1}\cdots B_{j+r-1}A_{j+r}\rangle\\
&-\left(\langle\eta_{z_{1}}B_{N}\rangle\langle A_{j+2}\eta_{z_{1}}^{\dagger}\rangle-\langle\eta_{z_{1}}A_{j+2}\rangle\langle B_{N}\eta_{z_{1}}^{\dagger}\rangle\right) \langle A_{N}B_{j}A_{j+1}B_{j+1}B_{j+3}\cdots B_{j+r-1}A_{j+r}\rangle\\
&-\cdots\cdots.
\end{aligned}
\end{equation}
Then we can write it into a $(r+1)$-th order determinant,
\begin{equation}
T_{j,j+r,N}=-\det\left[  \begin{array}{cccccc}
G_{N,N}-F_{N,N}&G_{N,j+1}-F_{N,j+1}&\cdots&G_{N,j+r}-F_{N,j+r}\\
G_{j,N}-F_{j,N}&G_{j,j+1}-F_{j,j+1}&\cdots&G_{j,j+r}-F_{j,j+r}\\
G_{j+1,N}-F_{j+1,N}&G_{j+1,j+1}-F_{j+1,j+1}&\cdots&G_{j+1,j+r}-F_{j+1,j+r}\\
&\cdots\cdots&&\\
G_{j+r-1,N}-F_{j+r-1,N}&G_{j+r-1,j+1}-F_{j+r-1,j+1}&\cdots&G_{j+r-1,j+r}-F_{j+r-1,j+r}\\
\end{array}\right] . \label{app_Tzxx}
\end{equation}.

\end{widetext}

\end{document}